\documentclass[
preprint,
 amsmath,amssymb,
 aps,
 prb,
superscriptaddress]{revtex4-2}
\usepackage{hyperref}
\hypersetup{
    colorlinks=true,
    linkcolor=blue,
    urlcolor=blue,
    citecolor=blue
}
\usepackage{standalone}
\usepackage{orcidlink}
\usepackage{graphicx}
\usepackage{dcolumn}
\usepackage{bm}
\usepackage[skip=0pt, indent=10pt]{parskip}
\usepackage{gensymb}
\usepackage{textgreek}
\usepackage{bm}
\usepackage{bbm}
\usepackage{xcolor}

\newcommand{\red}{\color{black}}

\newcommand{\mw}[1]{{\color{black}#1}}

\begin{document}

\title{Element-selective probing of ultrafast ferromagnetic--antiferromagnetic order dynamics in Fe/CoO bilayers}
\author{Chowdhury S. Awsaf}
\affiliation{Institut f\"ur Experimentalphysik, Freie Universit\"at Berlin, Arnimallee 14, 14195 Berlin, Germany}

\author{Sangeeta Thakur\orcidlink{0000-0003-4879-5650}}
\affiliation{Institut f\"ur Experimentalphysik, Freie Universit\"at Berlin, Arnimallee 14, 14195 Berlin, Germany}

\author{Markus Wei{\ss}enhofer\orcidlink{0000-0002-3283-2560}}
\affiliation{Institut f\"ur Experimentalphysik, Freie Universit\"at Berlin, Arnimallee 14, 14195 Berlin, Germany}
\affiliation{Department of Physics and Astronomy, Uppsala University, Box 516, 75120 Uppsala, Sweden}

\author{Jendrik G\"ordes\orcidlink{0000-0003-4321-8133}}
\affiliation{Institut f\"ur Experimentalphysik, Freie Universit\"at Berlin, Arnimallee 14, 14195 Berlin, Germany}

\author{Marcel Walter}
\affiliation{Institut f\"ur Experimentalphysik, Freie Universit\"at Berlin, Arnimallee 14, 14195 Berlin, Germany}

\author{Niko Pontius\orcidlink{0000-0002-5658-1751}}
\affiliation{Helmholtz-Zentrum Berlin f\"ur Materialien und Energie, Albert-Einstein-Stra{\ss}e 15, 12489 Berlin, Germany}

\author{Christian Sch\"u{\ss}ler-Langeheine\orcidlink{0000-0002-4553-9726}}
\affiliation{Helmholtz-Zentrum Berlin f\"ur Materialien und Energie, Albert-Einstein-Stra{\ss}e 15, 12489 Berlin, Germany}

\author{Peter M. Oppeneer\orcidlink{0000-0002-9069-2631}}
\affiliation{Department of Physics and Astronomy, Uppsala University, Box 516, 75120 Uppsala, Sweden}

\author{Wolfgang Kuch\orcidlink{0000-0002-5764-4574}}
\email{Correspondence and requests for materials should be addressed to W.K. (email: kuch@physik.fu-berlin.de)}
\affiliation{Institut f\"ur Experimentalphysik, Freie Universit\"at Berlin, Arnimallee 14, 14195 Berlin, Germany}

\date{\today}

\begin{abstract}
The ultrafast magnetization dynamics of an epitaxial Fe/CoO bilayer on Ag(001) is examined in an element-resolved way by resonant soft-x-ray reflectivity.  The transient magnetic linear dichroism at the Co $L_2$ edge and the magnetic circular dichroism at the Fe $L_3$ edge measured in reflection in a pump--probe experiment with 120 fs temporal resolution show the loss of antiferromagnetic and ferromagnetic order in CoO and Fe, respectively, both within 300 fs after excitation with 60 fs light pulses of 800 and 400 nm wavelengths.  Comparison to spin-dynamics simulations using an atomistic spin model shows that direct energy transfer from the laser-excited electrons in Fe to the magnetic moments in CoO 
provides the dominant
demagnetization channel in the case of 800-nm excitation.
\end{abstract}

\maketitle



The 
discovery of ultrafast magnetization dynamics research by Beaurepaire \textit{et al.}\ \cite{beaurepaire1996ultrafast} in ferromagnetic (FM) Ni has propelled scientists to investigate emergent ultrafast magnetic phenomena 
and theorize their possible explanations \cite{koopmans2010explaining, battiato2010superdiffusive, carpene2008dynamics, krauss2009ultrafast, dewhurst2018laser, holldack2010ultrafast, RevModPhys.82.2731, kirilyuk2013laser, walowski2016perspective, buzzi2018probing, CARVA2017291, wang2020ultrafast, SCHEID2022169596}. In recent times, antiferromagnetic (AFM) films have shown great promise for potential applications in memory \cite{marti2014room, wadley2016electrical, kriegner2016multiple, olejnik2017antiferromagnetic} and ultrafast spintronic devices \cite{jungwirth2016antiferromagnetic, baltz2018antiferromagnetic, nvemec2018antiferromagnetic, xiong2022antiferromagnetic}. Antiparallel sublattice magnetization eliminates the necessity for overall angular momentum dissipation, resulting in faster magnetization dynamics \cite{rettig2016itinerant, thielemann2017ultrafast}.
Moreover, AFM's can be used to tune
the magnetic properties of adjacent ferromagnetic films in FM/AFM hybrid layers, for example to approach specific functionalities. 
In the ultrafast regime, FM/AFM systems are interesting from a fundamental viewpoint with respect to the interplay of the different local and nonlocal mechanisms \cite{kumberg2020accelerating,1854} as well as the magnetic coupling at the interface that may govern the ultrafast response of the system \cite{eschenlohr2020spin}.
While the static interaction between FM and AFM layers has been extensively studied in several systems in the past, little is known about the ultrafast dynamic response of FM/AFM layered systems, despite the importance of interface effects \cite{RevModPhys.89.025006}.
This is mainly due to the zero net magnetic moment of AFM's, which hampers investigations in general, but time-resolved studies in particular since it excludes commonly-used methods detecting the ultrafast temporal evolution of the total magnetic moment. To address the question how and on which timescales the optical excitation of an FM/AFM bilayer is transferred between the two layers and into the magnetic subsystems, the temporal evolution of both the FM and AFM magnetic order has to be traced on ultrafast timescales.  Exciting both layers simultaneously or only one of them provides additional information on the interplay between different transfer paths.

Magnetic linear dichroism, which scales quadratically with the sublattice magnetization$M$, can be used to characterize collinear AFM order.  Although typically being much smaller than its circular counterpart,
it has been used in the visible-light regime 
to study 
the temporal evolution of spin order in CuMnAs or CoO films \cite{saidl2017optical, zheng2018magneto}.  
For the investigation of AFM/FM layered systems, however, elemental specificity is mandatory to separate the signals from the two layers, which is not provided by visible light.  Elemental specificity, on the other hand, is routinely achieved in the soft-x-ray regime, where one takes advantage of the differences of elemental absorption energies.  Time-resolved x-ray-spectroscopic studies on element-specific AFM order so far have utilized femtosecond soft-x-ray resonant diffraction \cite{holldack2010ultrafast, rettig2016itinerant, thielemann2017ultrafast, buzzi2018probing}.  Thereby, a relatively large unit cell of the AFM order is required, comparable to the wavelength of the incident x rays.
This does not work for AFM materials with simpler spin structures, where the size of the magnetic unit cell is 
only twice the size of the structural one.  

In this study, we employ time-resolved x-ray magnetic linear dichroism 
in resonant soft-x-ray reflectivity (R-XMLD) \cite{Oppeneer2003} to observe the ultrafast dynamics of AFM spins in a single-crystalline CoO film in an Fe/CoO bilayer upon excitation by pump pulses of 800 nm or 400 nm wavelength, with photon energies below and above the band gap of CoO of $\approx 2.5$ eV \cite{ma2015ultrafast}, respectively.  We detect the element-resolved dynamic response of the AFM CoO layer and  juxtapose it with the demagnetization in the adjacent FM Fe layer, obtained from time-resolved x-ray magnetic circular dichroism 
 in reflection (R-XMCD) \cite{Mertins2002}, to reveal the
dynamic behavior of the FM/AFM layered system. Interestingly, both layers demonstrate an ultrafast reduction of magnetic order with similar time constants of about 200--400 fs at both pump wavelengths. At the 800 nm pump, the excitation in the CoO layer must be entirely transferred from the Fe layer, as the 800 nm pump photon energy of 1.55 eV is smaller than the CoO bandgap, while at the 400 nm pump, a comparison of the demagnetization amplitudes of both layers shows a significant excitation directly in the CoO layer. We compare the experimental results to atomistic spin-dynamics
simulations using the stochastic Landau-Lifshitz-Gilbert (LLG) equation and a temperature model for the different layers and identify the relevant mechanisms governing the ultrafast spin dynamics in the FM/AFM bilayer.  


A film of 9 ($\pm$ 0.5) atomic monolayers (ML) of CoO is grown epitaxially on a Ag(001) surface, following the recipe described in Ref.\ \cite{abrudan2008structural}, and capped by 9 ($\pm$ 1) ML Fe
(for details of the sample preparation, see the Supplemental Material (SM) \cite{supp}).  CoO in Fe/CoO/Ag(001) has a collinear antiferromagnetic spin structure below about 290 K, aligned with the Fe magnetization direction along an 
 Fe $<$100$>$ easy axis of magnetization due to a strong coupling at the interface, and exhibits a characteristic XMLD in absorption upon turning the polarization axis of linearly polarized x rays of normal incidence by $90^{\circ}$
 \cite{abrudan2008structural, miguel2009magnetic}.  
 The AFM spin axis can thus be turned by $90^{\circ}$ in the sample plane by an external magnetic field via the coupling to the Fe magnetization.  

\begin{figure}
\includegraphics[width=0.75\columnwidth]{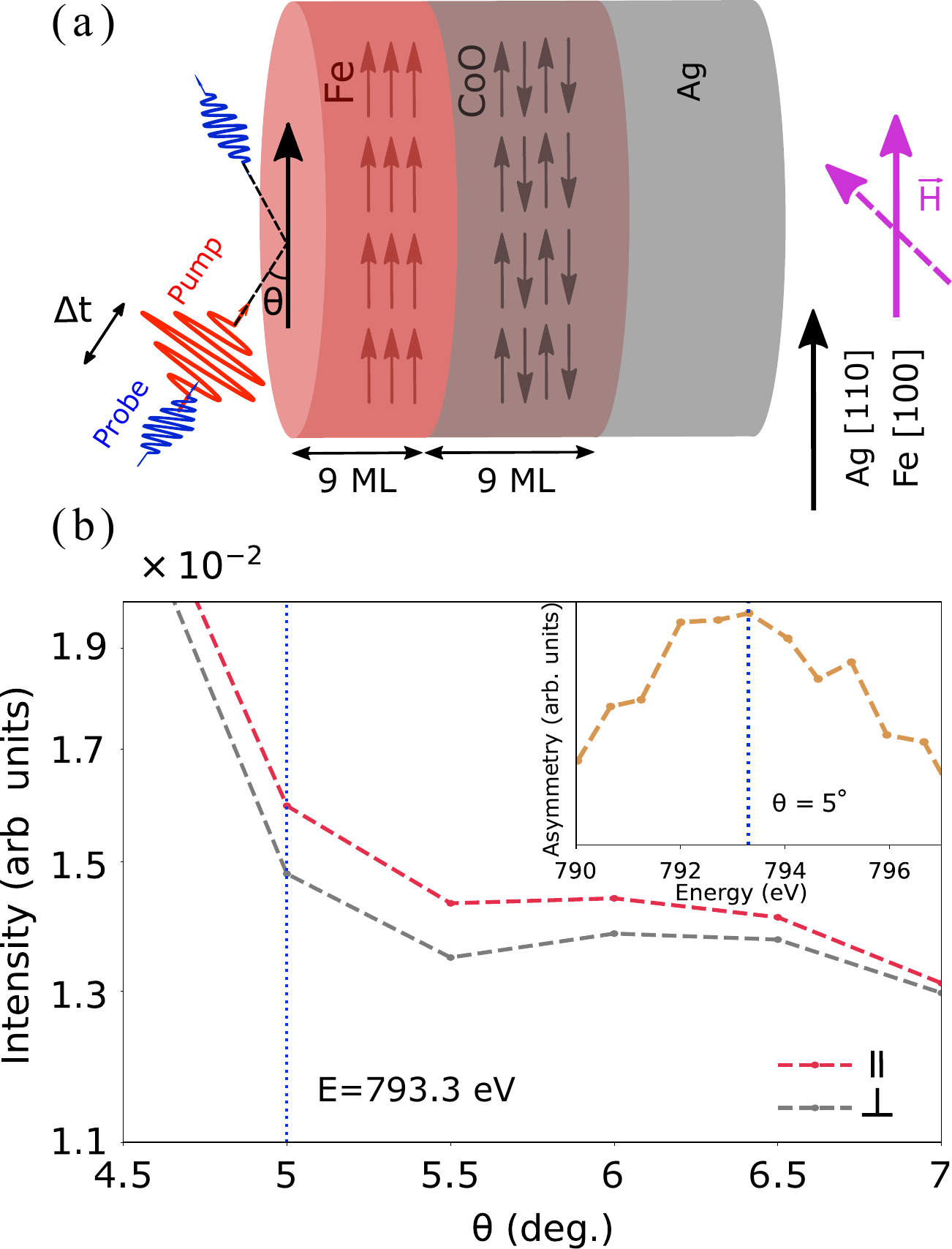}
\vspace*{-0.2cm}
\caption{\label{XMCD_XMLD_Fe_CoO}(a) Sketch of the sample and the experiment.  AFM order is probed with linearly $s$-polarized x rays tuned to the $L_2$ resonance of Co, co-propagating with a pump pulse, in reflectivity. The magnetic field $\emph{H}$ is changed by 90$^\circ$ along the sample surface to obtain R-XMLD contrast. (b) Reflected intensity for the parallel and perpendicular directions of the magnetic field as a function of reflection angle at the optimal photon energy for CoO, as established by comparing such scans for different photon energies.  The used incidence angle of $\theta = 5^{\circ}$ is marked by a vertical dashed line.  Inset:  Magnetic contrast as a function of photon energy at the constant angle of $\theta = 5^{\circ}$. For CoO, the least acquisition time at this angle is attained when the photon energy is 793.3 eV (blue vertical line).}
\end{figure}

The sample is transferred under ultra-high vacuum conditions to the synchrotron radiation source BESSY II in Berlin.  There, the time-resolved measurements are performed in 10$^{-8}$ mbar pressure, at 200 K with 120 mT applied field parallel to the sample surface for magnetic saturation, at the Femtoslicing Facility.
60 fs $p$-polarized pulses of either 800 or 400 nm wavelength are used to pump the sample at a repetition rate of 3 kHz, while subsequent probing is achieved with 100 fs polarized x-ray pulses from the femtoslicing mode of the beamline, probing the sample at 6 kHz, to detect both pumped and unpumped reflected signals alternatingly. The latter are used to normalize the pumped signal. Both pump and probe beams are co-propagating ($1^{\circ}$--$2^{\circ}$ apart to filter the pump pulse after reflection) onto the sample at grazing incidence with the pump spot size spanning approximately five times the probe spot size (100 $\mu$m) to excite the probed area uniformly. For R-XMLD of CoO and R-XMCD of Fe, we opted for 793.3 eV at the Co \textit{L}\textsubscript{2} edge and 709.8 eV at the Fe \textit{L}\textsubscript{3} edge while keeping the same 5$^{\circ}$ incident grazing angle in both cases in order to maintain identical pump conditions.  These energies were chosen to achieve the most efficient measurement condition considering acquisition times for the given angle of incidence \cite{supp}.  Figure \ref{XMCD_XMLD_Fe_CoO} shows in (a) a sketch of the sample and the experimental configuration and in (b) the static magnetic signal at the Co $L_2$ edge. 
The x-ray beam was maintained at linear $s$ polarization for the R-XMLD experiment, while the magnetic field was altered by means of the superconducting vector magnet of the beamline in steps of $90^{\circ}$ in the sample plane 
in order to change the magnetic axis of CoO between parallel and perpendicular to the x-ray polarization via the magnetization of the Fe layer.  This way, no structural linear dichroism contributes to the difference signal. The reflectivity change of about 5\% is thus entirely due to R-XMLD.  For the R-XMCD measurement of Fe, the incident x rays were circularly polarized and the magnetic field direction was reversed by $180^{\circ}$ between parallel and antiparallel to x-ray helicity vector.  

\begin{figure}[b]
\includegraphics[width=0.98\columnwidth]{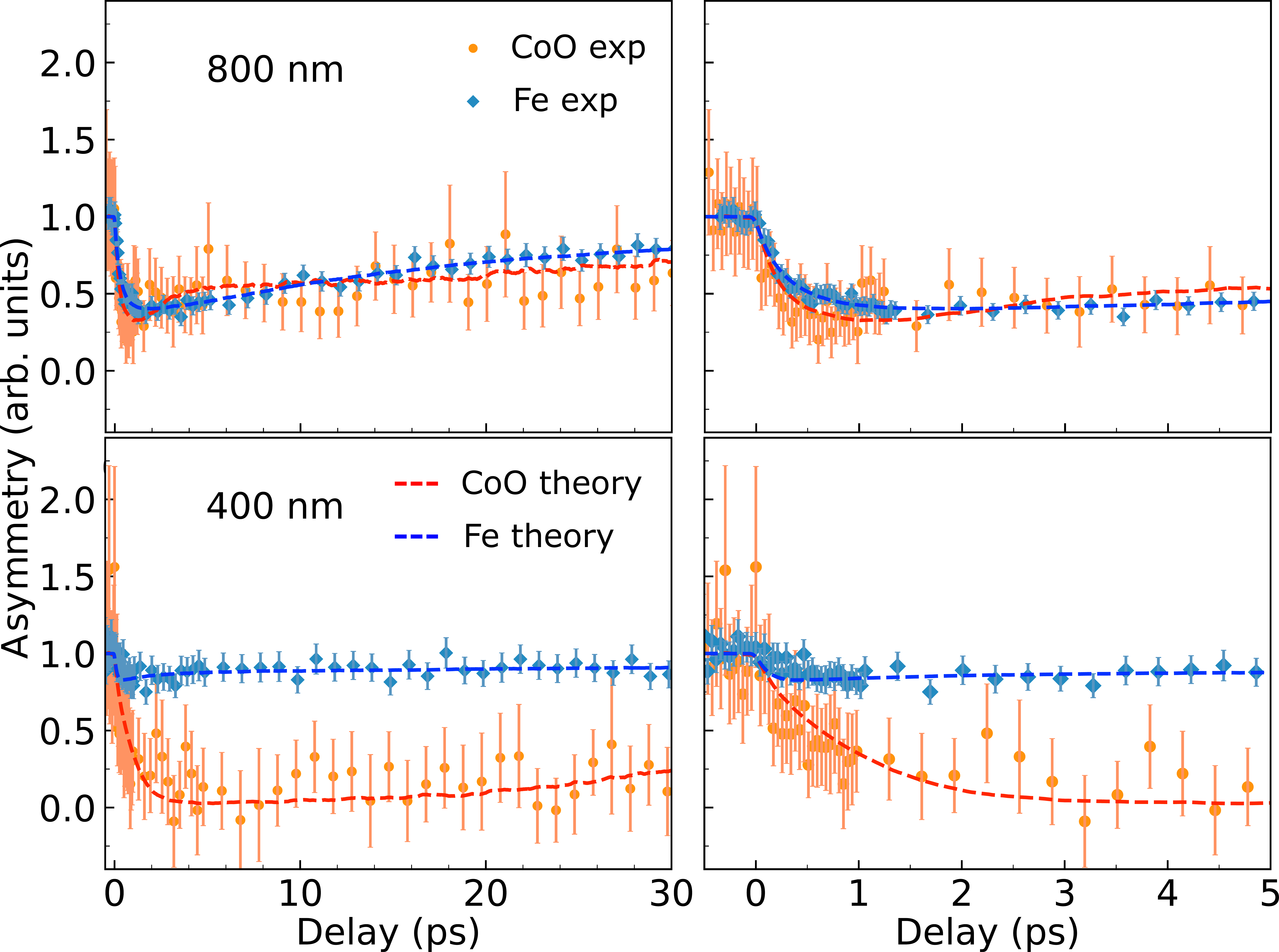}
\caption{\label{Fe_CoO_sim}Magnetization dynamics of CoO and Fe experimentally observed for 800 and 400 nm pump and presented as scans of the R-XMLD and R-XMCD asymmetry as a function of delay time. Both short- and long-range delay-time graphs (right and left panels, respectively) illustrate the different magnetization regimes. The scattered points are the experimental data, and the dashed lines result from a simulation using the atomistic spin model described in the text and schematically depicted in Fig.\ \ref{subsystem}.}
\end{figure}

Although the resulting magnetic signal in R-XMLD is relatively small and the sliced synchrotron-radiation probe exhibits about eight orders of magnitude reduced intensity compared to static experiments, it is possible to measure the time evolution of the magnetic signals from Fe and CoO layers at 200 K when pumped at 800 and 400 nm wavelength with 10 mJ/cm$^2$  incident fluence, as shown in Figure \ref{Fe_CoO_sim}.
Both Fe and CoO undergo a similar drop in magnetic asymmetry to around 30\% upon 800 nm excitation, while Fe demagnetizes less at 400 nm pump.  
For a quantitative evaluation, the magnetization dynamics were fitted with a double exponential function convoluted with a 120 fs Gaussian response function in order to describe the fast demagnetization and slower remagnetization of the experimental data within the measured time window \cite{supp}.   The resulting parameters, in the case of CoO taking into account the proportionality of the R-XMLD signal to the square of the sublattice magnetization, are summarized in Tab.\ \ref{parameters}.  The demagnetization times are somewhat shorter at 800 nm pump with 200--300 fs compared to 400 nm pump, where 300--450 fs are measured.  
The demagnetization amplitude of CoO is comparable at 800 and 400 nm, while that of Fe is significantly smaller at 400 nm.  

    \begin{table}[t!]
        \begin{adjustbox}{scale = 0.82,center}
        \begin{tabular}{|c|c|c|c|c|c|}
            \hline 
            Wavelength / nm & Layer & $\tau _{de}$ / fs & $A_{de}$ & $\tau _{re}$ / ps & $A_{re}$ \\
            \hline
            800 & CoO & $276 \pm 47$ &$0.485 \pm 0.047$ & $3.7 \pm 1.6$ &$0.197 \pm 0.043$ \\
            \hline
             800 & Fe & $213 \pm 16$  &$0.623 \pm 0.007$& $> 20$ &$0.51 \pm 0.03$  \\
            \hline
            400 & CoO & $450 \pm 130$ &$0.54 \pm 0.02$ & -- & 0\\
            \hline
             400 & Fe & $302 \pm 135$ &$0.18 \pm 0.02$ & $5.3 \pm 3.9$ &$0.08 \pm 0.02$  \\
            \hline
        \end{tabular}
        \end{adjustbox}
        \caption{Demagnetization and remagnetization times ($\tau_{de}$ and $\tau_{re}$) and amplitudes ($A_{de}$ and $A_{re}$) extracted from the experimental data by exponential fits for the CoO sublattice and the Fe transient magnetization at 800 and 400 nm pump wavelengths.}
        \label{parameters}
    \end{table}
    

800 nm photons are below the bandgap of CoO of approximately 2.5 eV \cite{ma2015ultrafast}.  In a standalone CoO layer, Zheng {\it et al.}\ consequently did not observe any response in the time-resolved Voigt-effect signal after 800 nm excitation \cite{zheng2018magneto}, due to CoO's transparency at this wavelength. Our experimental results for 800 nm excitation, therefore, suggest significant energy transfer to the CoO layer from the other layers, 
specifically from the Fe layer. This is consistent with previous reports from FM/AFM layered systems.  In a work by Ma {\it et al.}\/, a CoO/Fe bilayer showed an enhanced precession of the magnetization upon 800 nm excitation compared to a single Fe layer, as detected by time-resolved magneto-optical Kerr effect, which was interpreted as a modulation of the exchange anisotropy between Fe and CoO induced by the pump-generated hot electrons in Fe \cite{ma2015ultrafast}. Wust {\it et al.}\/  
have previously reported comparable findings for a below-bandgap excitation of Pt/NiO bilayers \cite{wust2022indirect}.  According to their findings, NiO can effectively demagnetize by an 800 nm laser pump when coated with a Pt layer. 

To describe the demagnetization of the Fe/CoO bilayer and the flow of energy between the layers, we model the magnetization dynamics using an atomistic spin model 
based on
stochastic Landau-Lifshitz-Gilbert equations for the spin degrees of freedom coupled to the respective electron and phonon temperatures via Gilbert damping parameters in both layers \cite{Nowak2007,Kazantseva2008}. The Heisenberg Hamiltonian for the spin degrees of freedom is
\begin{small}
\begin{align}\label{equ Hamiltonian}
            \mathcal{H}= -\sum_{ij} J_{ij} \bm{{S_{i}}} \cdot \bm{{S_{j}}} \, , 
\end{align}
\end{small}
where $J_{ij}$ are the coupling constants and $\bm{S}_{i}$ are unit vectors along the direction of the magnetic moments of Fe or Co atoms at lattice site $i$. For simplicity, we treat the system as simple cubic, assume perfect stacking at the interface and restrict the coupling to nearest neighbors,
with the coupling constants 
for CoO and Fe expressed as $J_{ij}=\frac{k_\mathrm{B}T_\mathrm{c}}{1.44}$ \cite{Garanin1996}, where $T_\mathrm{c} = 293$ K  and $1043$ K are used as N\'eel and Curie temperatures for CoO and Fe, respectively.  The interfacial coupling $J^\mathrm{if}$ is taken as a free parameter. The 
dynamics of the magnetic moments are described using the
stochastic 
LLG equation 
\cite{Landau1935,brown1963thermal,Gilbert2004},
\begin{small}
\begin{align}
            & \dot{\bm{S}_i}=-\frac{\gamma_i}{\mu_i}\bm{S}_i \times (\bm{H}_{i}+\bm{\zeta}_i)+(\alpha_i^\mathrm{e} + \alpha_i^\mathrm{ph})\bm{S}_i \times \dot{\bm{S}_i} , \label{equ LLG} \\
            & \langle \bm{\zeta}_i(t)\bm{\zeta}_i^T(t') \rangle = 2\frac{\mu_i}{\gamma_i}k_B(\alpha_i^\mathrm{e} T_i^\mathrm{e} + \alpha_i^\mathrm{ph} T_i^\mathrm{ph})\mathbbm{1} \delta_{ij}\delta(t-t') .\label{equ noise}
\end{align}
\end{small}
The first right-hand term in Eq.\ (\ref{equ LLG}) is the precession torque with gyromagnetic ratio $\gamma_i$, spin magnetic moment $\mu_i$, and $\bm{\zeta}_i(t)$ representing thermal fluctuations to the effective field $\bm{H}_{i}=-\partial \mathcal{H}/\partial \bm{S}_{i}$ \cite{brown1963thermal}. The second term is the damping torque, where $\alpha_i^\mathrm{e}$ and $\alpha_i^\mathrm{ph}$ are the Gilbert damping constants that couple the respective spin to the electron and phonon subsystems. We assumed a homogeneous distribution of both, electron and phonon temperatures, in each layer. 
Furthermore, we took $\alpha_\mathrm{Fe}^\mathrm{ph} = 0$, which is reasonable for 3$d$ transition metals \cite{Zahn2021,Zahn2022}. 

\begin{figure}[b]
\includegraphics[width=0.98\columnwidth]{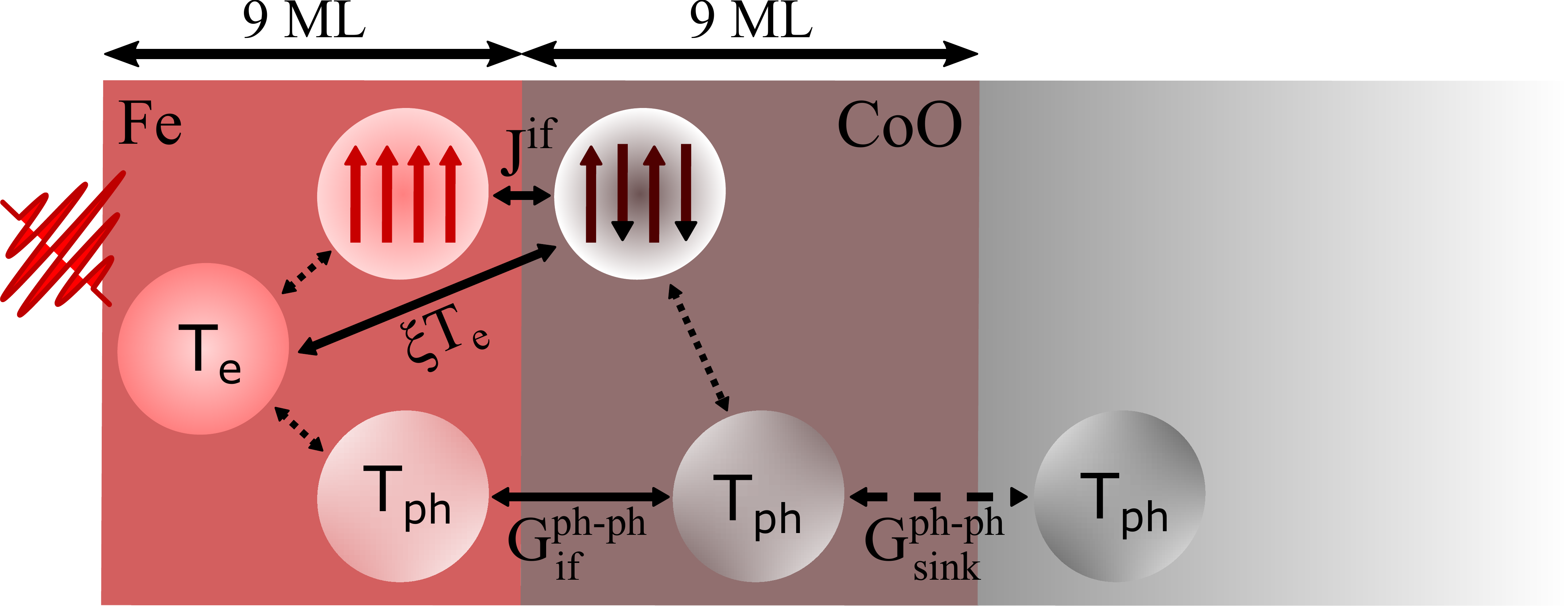}
\vspace*{-0.1cm}
\caption{\label{subsystem}Representation of different subsystems of the Fe/CoO bilayer on the Ag substrate (see text). The double-headed arrows symbolize the interactions between the different subsystems considered in the model with the solid ones describing the three energy transfer channels from Fe to CoO.}
\end{figure}

An energy-conserving temperature model is employed that records the temperature evolution of the electrons ($T_\mathrm{e}$) in Fe and the phonons ($T_\mathrm{ph}$) in Fe and CoO, see Fig.\
\ref{subsystem} and the SM
\cite{supp}. Three energy transfer channels from the Fe to the CoO layer with their respective parameters are considered, as schematically depicted in Fig.\ \ref{subsystem}: the interfacial exchange interaction $J^\mathrm{if}$ between Fe and CoO spins, 
the interfacial heat transfer coefficient that couples phonons between Fe and CoO, $G_\mathrm{if}^\mathrm{ph-ph}$
, and, in addition, we introduce a third energy transfer channel by assuming that at the interface, the CoO spins interact with the Fe electron subsystem, similar to Ref.~\cite{wust2022indirect}. Within our simulations, this energy transfer, which could stem e.g. from $s-d$ coupling at the interface \cite{Chen2015}, was modeled as incoherent by using the temperature of the Fe electrons in the evaluation of the thermal fluctuations for the Co magnetic moments via Eq.~\eqref{equ noise}, with an associated damping coefficient $\alpha_\mathrm{Fe\rightarrow CoO}^\mathrm{e} = \xi T_\mathrm{Fe}^\mathrm{e}$ (see SM \cite{supp} for details.) In doing so, 
the linear scaling of $\alpha_\mathrm{Fe\rightarrow CoO}^\mathrm{e}$ with temperature means that the coupling peaks strongly during the laser pulse (when the electronic temperature reaches thousands of Kelvin).

The simulations reveal that phononic and magnonic contributions result only in relatively slow demagnetization of the CoO sublattice magnetization of several picoseconds, and only the electronic channel leads to demagnetization at subpicosecond timescale \cite{supp}.  This shows that direct Fe electron--CoO spin coupling plays the dominant role in quenching the CoO antiferromagnetic order when pumped with 800 nm wavelength.  
By tuning the free parameters $\alpha_\mathrm{Fe}^\mathrm{e}$, $J^\mathrm{if}$, $\alpha_\mathrm{CoO}^\mathrm{ph}$, 
\mw{$\xi$}, $G_\mathrm{if}^\mathrm{ph-ph}$ and $G_\mathrm{sink}^\mathrm{ph-ph}$, the model can reproduce both the R-XMLD and R-XMCD data simultaneously as depicted in Fig.\ \ref{Fe_CoO_sim} by dashed lines.  Table S II of the SM \cite{supp} presents the parameters used to create the lines in Fig.\ \ref{Fe_CoO_sim}.  

Implementing the same model, in which the CoO layer is exclusively excited via energy transfer from the Fe layer, to the 400 nm pump data resulted in significantly lower demagnetization of CoO than experimentally observed. Because of the relatively small demagnetization amplitude of the Fe layer at 400 nm, see Fig.\ \ref{Fe_CoO_sim}, the Co R-XMLD signal would then reduce only by 10--15\% compared to the observed 80\% \cite{supp}. We therefore have to consider a direct excitation of the CoO layer by 400 nm photons, corresponding to 3.1 eV photon energy, which is responsible for the major part of the quench of AFM order in CoO when pumped at 400 nm. At that wavelength, photons can excite electrons from O 2\textit{p} to Co 3\textit{d} states and spin transfer can occur between adjacent Co sites due to the AFM alignment \cite{zheng2018magneto, ma2015ultrafast}.  

It is {\it a priori}\/ not clear how to include such a direct excitation of CoO into our model.  One possibility, which we implement here to demonstrate that the experimental data is consistent with the excitation of the CoO layer, is to include an electron temperature of CoO and a finite coupling of CoO spins to CoO electrons, $\alpha_\mathrm{CoO}^\mathrm{e}$, into the model described above (see \cite{supp}). 
We use linearized electron heat capacities $C_\mathrm{CoO}^\mathrm{e}$ = $\gamma_\mathrm{CoO}^\mathrm{e} T_\mathrm{CoO}^\mathrm{e}$, although the Sommerfeld model of linear electron heat capacity for metals may be an overestimation for CoO, and a constant (temperature-independent) damping parameter $\alpha_\mathrm{CoO}^\mathrm{e}$.
By tuning the relevant parameters, we can replicate the experimental results, as demonstrated in Fig.\ \ref{Fe_CoO_sim}.  The parameters are presented in the SM
\cite{supp}.  
We emphasize that this is only one out of several possible scenarios to model the photoexcitation of CoO, but it shows that such a direct excitation of CoO by the laser pulse explains the experimental result, in particular the larger demagnetization amplitude of CoO compared to that of Fe.

In principle also spin transport between the layers could be discussed as a possibility to explain the different demagnetization amplitudes of CoO and Fe at 400 nm pump. Kumberg {\it et al.}\ \cite{kumberg2020accelerating} suggest that magnetic FM/AFM stacks can facilitate the entry of minority spin currents into the FM, leading to its faster demagnetization. Conversely, if the FM layer's spins are collinear to the ones in the AFM, more majority spins can enter the AFM layer than for a noncollinear alignment.  However, in that experiment, only the demagnetization time was affected by the presence of AFM order in the adjacent layer, with no effect on the amplitudes. 
The electron excitation at 3.1 eV is above the bandgap in CoO.  So, spin injection from CoO to Fe may only occur at high energies. Since there is a substantial unoccupied minority $d$ density between 0 and 3 eV in Fe, more spin-minority electrons can enter the Fe layer and increase the Fe demagnetization.  However, this is in contrast to the observed smaller Fe demagnetization (Fig.\ \ref{Fe_CoO_sim}) at 3.1 eV pump.  For this, one would need to assume a spin current with a preference for transferring minority electrons from Fe into the CoO or spin-polarized electrons from CoO to Fe with Fe majority spin.  We thus expect that spin transport between the layers does not play a major role here.

To conclude, using time-resolved XMLD in soft-x-ray reflectivity with 120 fs temporal resolution, we 
observed the ultrafast spin dynamics in CoO after excitation of an Fe/CoO bilayer by 60 fs pulses of 800 and 400 nm wavelengths on an element-resolved basis.  This allows to compare it with the corresponding element- and time-resolved R-XMCD of the Fe layer to investigate the ultrafast AFM and FM demagnetizations and the interfacial transfer of excitation between the AFM CoO and the FM Fe layer. 
We find that CoO AFM and Fe FM orders demagnetize similarly fast.
Our atomistic spin-dynamics
model with stochastic LLG shows that energy transfer via the coupling of hot Fe electrons to CoO spins is the primary mechanism for the rapid quenching of CoO magnetic moments on the order of 300 fs for excitation below the band gap of CoO.  
The loss of AFM order in the CoO layer is thus entirely due to energy transfer from the interface, which spreads with the time constant of 276 fs through the entire 9-ML film of CoO.
In the case of above-bandgap excitation at 400 nm pump, the magnetic order in CoO reduces much more than the one in Fe, which can only be explained by considering the direct excitation of CoO electrons at that wavelength. Including this 
in the atomistic simulations 
describes the experimental observation at 400 nm pump. 
Our study further advocates that
the elemental resolution of R-XMLD makes it a 
promising option for time-resolved magnetization research 
of AFM's and AFM heterostructures.

\begin{acknowledgments}
		This work was supported by the Deutsche Forschungsgemeinschaft via the CRC/TRR 227 ``Ultrafast Spin Dynamics", {\red project-ID: 328545488,} projects A03, A07, and Z.  We further acknowledge support from the K.\ and A.\ Wallenberg Foundation (Grants No.\ 2022.0079 and 2023.0336). We thank the Helmholtz-Zentrum Berlin for the allocation of synchrotron radiation beamtime, I.\ Gelen for help with the sample preparation chamber, and M.\ A.\ Mawass for assistance during the beamtime. Computational resources were provided by the National Academic Infrastructure for Supercomputing in Sweden (NAISS) at NSC Link\"oping partially funded by the Swedish Research Council through grant agreement No.\ 2022-06725.
\end{acknowledgments}


\begin{thebibliography}{46}%
\makeatletter
\providecommand \@ifxundefined [1]{%
 \@ifx{#1\undefined}
}%
\providecommand \@ifnum [1]{%
 \ifnum #1\expandafter \@firstoftwo
 \else \expandafter \@secondoftwo
 \fi
}%
\providecommand \@ifx [1]{%
 \ifx #1\expandafter \@firstoftwo
 \else \expandafter \@secondoftwo
 \fi
}%
\providecommand \natexlab [1]{#1}%
\providecommand \enquote  [1]{``#1''}%
\providecommand \bibnamefont  [1]{#1}%
\providecommand \bibfnamefont [1]{#1}%
\providecommand \citenamefont [1]{#1}%
\providecommand \href@noop [0]{\@secondoftwo}%
\providecommand \href [0]{\begingroup \@sanitize@url \@href}%
\providecommand \@href[1]{\@@startlink{#1}\@@href}%
\providecommand \@@href[1]{\endgroup#1\@@endlink}%
\providecommand \@sanitize@url [0]{\catcode `\\12\catcode `\$12\catcode
  `\&12\catcode `\#12\catcode `\^12\catcode `\_12\catcode `\%12\relax}%
\providecommand \@@startlink[1]{}%
\providecommand \@@endlink[0]{}%
\providecommand \url  [0]{\begingroup\@sanitize@url \@url }%
\providecommand \@url [1]{\endgroup\@href {#1}{\urlprefix }}%
\providecommand \urlprefix  [0]{URL }%
\providecommand \Eprint [0]{\href }%
\providecommand \doibase [0]{https://doi.org/}%
\providecommand \selectlanguage [0]{\@gobble}%
\providecommand \bibinfo  [0]{\@secondoftwo}%
\providecommand \bibfield  [0]{\@secondoftwo}%
\providecommand \translation [1]{[#1]}%
\providecommand \BibitemOpen [0]{}%
\providecommand \bibitemStop [0]{}%
\providecommand \bibitemNoStop [0]{.\EOS\space}%
\providecommand \EOS [0]{\spacefactor3000\relax}%
\providecommand \BibitemShut  [1]{\csname bibitem#1\endcsname}%
\let\auto@bib@innerbib\@empty
\bibitem [{\citenamefont {Beaurepaire}\ \emph {et~al.}(1996)\citenamefont
  {Beaurepaire}, \citenamefont {Merle}, \citenamefont {Daunois},\ and\
  \citenamefont {Bigot}}]{beaurepaire1996ultrafast}%
  \BibitemOpen
  \bibfield  {author} {\bibinfo {author} {\bibfnamefont {E.}~\bibnamefont
  {Beaurepaire}}, \bibinfo {author} {\bibfnamefont {J.-C.}\ \bibnamefont
  {Merle}}, \bibinfo {author} {\bibfnamefont {A.}~\bibnamefont {Daunois}},\
  and\ \bibinfo {author} {\bibfnamefont {J.-Y.}\ \bibnamefont {Bigot}},\
  }\bibfield  {title} {\bibinfo {title} {Ultrafast spin dynamics in
  ferromagnetic nickel},\ }\href {https://doi.org/10.1103/PhysRevLett.76.4250}
  {\bibfield  {journal} {\bibinfo  {journal} {Phys. Rev. Lett.}\ }\textbf
  {\bibinfo {volume} {76}},\ \bibinfo {pages} {4250} (\bibinfo {year}
  {1996})}\BibitemShut {NoStop}%
\bibitem [{\citenamefont {Koopmans}\ \emph {et~al.}(2010)\citenamefont
  {Koopmans}, \citenamefont {Malinowski}, \citenamefont {Dalla~Longa},
  \citenamefont {Steiauf}, \citenamefont {F{\"a}hnle}, \citenamefont {Roth},
  \citenamefont {Cinchetti},\ and\ \citenamefont
  {Aeschlimann}}]{koopmans2010explaining}%
  \BibitemOpen
  \bibfield  {author} {\bibinfo {author} {\bibfnamefont {B.}~\bibnamefont
  {Koopmans}}, \bibinfo {author} {\bibfnamefont {G.}~\bibnamefont
  {Malinowski}}, \bibinfo {author} {\bibfnamefont {F.}~\bibnamefont
  {Dalla~Longa}}, \bibinfo {author} {\bibfnamefont {D.}~\bibnamefont
  {Steiauf}}, \bibinfo {author} {\bibfnamefont {M.}~\bibnamefont {F{\"a}hnle}},
  \bibinfo {author} {\bibfnamefont {T.}~\bibnamefont {Roth}}, \bibinfo {author}
  {\bibfnamefont {M.}~\bibnamefont {Cinchetti}},\ and\ \bibinfo {author}
  {\bibfnamefont {M.}~\bibnamefont {Aeschlimann}},\ }\bibfield  {title}
  {\bibinfo {title} {Explaining the paradoxical diversity of ultrafast
  laser-induced demagnetization},\ }\href {https://doi.org/10.1038/nmat2593}
  {\bibfield  {journal} {\bibinfo  {journal} {Nat. Mat.}\ }\textbf {\bibinfo
  {volume} {9}},\ \bibinfo {pages} {259} (\bibinfo {year} {2010})}\BibitemShut
  {NoStop}%
\bibitem [{\citenamefont {Battiato}\ \emph {et~al.}(2010)\citenamefont
  {Battiato}, \citenamefont {Carva},\ and\ \citenamefont
  {Oppeneer}}]{battiato2010superdiffusive}%
  \BibitemOpen
  \bibfield  {author} {\bibinfo {author} {\bibfnamefont {M.}~\bibnamefont
  {Battiato}}, \bibinfo {author} {\bibfnamefont {K.}~\bibnamefont {Carva}},\
  and\ \bibinfo {author} {\bibfnamefont {P.~M.}\ \bibnamefont {Oppeneer}},\
  }\bibfield  {title} {\bibinfo {title} {Superdiffusive spin transport as a
  mechanism of ultrafast demagnetization},\ }\href
  {https://doi.org/10.1103/PhysRevLett.105.027203} {\bibfield  {journal}
  {\bibinfo  {journal} {Phys. Rev. Lett.}\ }\textbf {\bibinfo {volume} {105}},\
  \bibinfo {pages} {027203} (\bibinfo {year} {2010})}\BibitemShut {NoStop}%
\bibitem [{\citenamefont {Carpene}\ \emph {et~al.}(2008)\citenamefont
  {Carpene}, \citenamefont {Mancini}, \citenamefont {Dallera}, \citenamefont
  {Brenna}, \citenamefont {Puppin},\ and\ \citenamefont
  {De~Silvestri}}]{carpene2008dynamics}%
  \BibitemOpen
  \bibfield  {author} {\bibinfo {author} {\bibfnamefont {E.}~\bibnamefont
  {Carpene}}, \bibinfo {author} {\bibfnamefont {E.}~\bibnamefont {Mancini}},
  \bibinfo {author} {\bibfnamefont {C.}~\bibnamefont {Dallera}}, \bibinfo
  {author} {\bibfnamefont {M.}~\bibnamefont {Brenna}}, \bibinfo {author}
  {\bibfnamefont {E.}~\bibnamefont {Puppin}},\ and\ \bibinfo {author}
  {\bibfnamefont {S.}~\bibnamefont {De~Silvestri}},\ }\bibfield  {title}
  {\bibinfo {title} {Dynamics of electron-magnon interaction and ultrafast
  demagnetization in thin iron films},\ }\href
  {https://doi.org/10.1103/PhysRevB.78.174422} {\bibfield  {journal} {\bibinfo
  {journal} {Phys. Rev. B}\ }\textbf {\bibinfo {volume} {78}},\ \bibinfo
  {pages} {174422} (\bibinfo {year} {2008})}\BibitemShut {NoStop}%
\bibitem [{\citenamefont {Krau\ss{}}\ \emph {et~al.}(2009)\citenamefont
  {Krau\ss{}}, \citenamefont {Roth}, \citenamefont {Alebrand}, \citenamefont
  {Steil}, \citenamefont {Cinchetti}, \citenamefont {Aeschlimann},\ and\
  \citenamefont {Schneider}}]{krauss2009ultrafast}%
  \BibitemOpen
  \bibfield  {author} {\bibinfo {author} {\bibfnamefont {M.}~\bibnamefont
  {Krau\ss{}}}, \bibinfo {author} {\bibfnamefont {T.}~\bibnamefont {Roth}},
  \bibinfo {author} {\bibfnamefont {S.}~\bibnamefont {Alebrand}}, \bibinfo
  {author} {\bibfnamefont {D.}~\bibnamefont {Steil}}, \bibinfo {author}
  {\bibfnamefont {M.}~\bibnamefont {Cinchetti}}, \bibinfo {author}
  {\bibfnamefont {M.}~\bibnamefont {Aeschlimann}},\ and\ \bibinfo {author}
  {\bibfnamefont {H.~C.}\ \bibnamefont {Schneider}},\ }\bibfield  {title}
  {\bibinfo {title} {Ultrafast demagnetization of ferromagnetic transition
  metals: The role of the {C}oulomb interaction},\ }\href
  {https://doi.org/10.1103/PhysRevB.80.180407} {\bibfield  {journal} {\bibinfo
  {journal} {Phys. Rev. B}\ }\textbf {\bibinfo {volume} {80}},\ \bibinfo
  {pages} {180407(R)} (\bibinfo {year} {2009})}\BibitemShut {NoStop}%
\bibitem [{\citenamefont {Dewhurst}\ \emph {et~al.}(2018)\citenamefont
  {Dewhurst}, \citenamefont {Elliott}, \citenamefont {Shallcross},
  \citenamefont {Gross},\ and\ \citenamefont {Sharma}}]{dewhurst2018laser}%
  \BibitemOpen
  \bibfield  {author} {\bibinfo {author} {\bibfnamefont {J.~K.}\ \bibnamefont
  {Dewhurst}}, \bibinfo {author} {\bibfnamefont {P.}~\bibnamefont {Elliott}},
  \bibinfo {author} {\bibfnamefont {S.}~\bibnamefont {Shallcross}}, \bibinfo
  {author} {\bibfnamefont {E.~K.~U.}\ \bibnamefont {Gross}},\ and\ \bibinfo
  {author} {\bibfnamefont {S.}~\bibnamefont {Sharma}},\ }\bibfield  {title}
  {\bibinfo {title} {Laser-induced intersite spin transfer},\ }\href
  {https://doi.org/10.1021/acs.nanolett.7b05118} {\bibfield  {journal}
  {\bibinfo  {journal} {Nano Lett}\ }\textbf {\bibinfo {volume} {18}},\
  \bibinfo {pages} {1842} (\bibinfo {year} {2018})}\BibitemShut {NoStop}%
\bibitem [{\citenamefont {Holldack}\ \emph {et~al.}(2010)\citenamefont
  {Holldack}, \citenamefont {Pontius}, \citenamefont {Schierle}, \citenamefont
  {Kachel}, \citenamefont {Soltwisch}, \citenamefont {Mitzner}, \citenamefont
  {Quast}, \citenamefont {Springholz},\ and\ \citenamefont
  {Weschke}}]{holldack2010ultrafast}%
  \BibitemOpen
  \bibfield  {author} {\bibinfo {author} {\bibfnamefont {K.}~\bibnamefont
  {Holldack}}, \bibinfo {author} {\bibfnamefont {N.}~\bibnamefont {Pontius}},
  \bibinfo {author} {\bibfnamefont {E.}~\bibnamefont {Schierle}}, \bibinfo
  {author} {\bibfnamefont {T.}~\bibnamefont {Kachel}}, \bibinfo {author}
  {\bibfnamefont {V.}~\bibnamefont {Soltwisch}}, \bibinfo {author}
  {\bibfnamefont {R.}~\bibnamefont {Mitzner}}, \bibinfo {author} {\bibfnamefont
  {T.}~\bibnamefont {Quast}}, \bibinfo {author} {\bibfnamefont
  {G.}~\bibnamefont {Springholz}},\ and\ \bibinfo {author} {\bibfnamefont
  {E.}~\bibnamefont {Weschke}},\ }\bibfield  {title} {\bibinfo {title}
  {{Ultrafast dynamics of antiferromagnetic order studied by femtosecond
  resonant soft x-ray diffraction}},\ }\href
  {https://doi.org/10.1063/1.3474612} {\bibfield  {journal} {\bibinfo
  {journal} {Appl. Phys. Lett.}\ }\textbf {\bibinfo {volume} {97}},\ \bibinfo
  {pages} {062502} (\bibinfo {year} {2010})}\BibitemShut {NoStop}%
\bibitem [{\citenamefont {Kirilyuk}\ \emph {et~al.}(2010)\citenamefont
  {Kirilyuk}, \citenamefont {Kimel},\ and\ \citenamefont
  {Rasing}}]{RevModPhys.82.2731}%
  \BibitemOpen
  \bibfield  {author} {\bibinfo {author} {\bibfnamefont {A.}~\bibnamefont
  {Kirilyuk}}, \bibinfo {author} {\bibfnamefont {A.~V.}\ \bibnamefont
  {Kimel}},\ and\ \bibinfo {author} {\bibfnamefont {T.}~\bibnamefont
  {Rasing}},\ }\bibfield  {title} {\bibinfo {title} {Ultrafast optical
  manipulation of magnetic order},\ }\href
  {https://doi.org/10.1103/RevModPhys.82.2731} {\bibfield  {journal} {\bibinfo
  {journal} {Rev. Mod. Phys.}\ }\textbf {\bibinfo {volume} {82}},\ \bibinfo
  {pages} {2731} (\bibinfo {year} {2010})}\BibitemShut {NoStop}%
\bibitem [{\citenamefont {Kirilyuk}\ \emph {et~al.}(2013)\citenamefont
  {Kirilyuk}, \citenamefont {Kimel},\ and\ \citenamefont
  {Rasing}}]{kirilyuk2013laser}%
  \BibitemOpen
  \bibfield  {author} {\bibinfo {author} {\bibfnamefont {A.}~\bibnamefont
  {Kirilyuk}}, \bibinfo {author} {\bibfnamefont {A.~V.}\ \bibnamefont
  {Kimel}},\ and\ \bibinfo {author} {\bibfnamefont {T.}~\bibnamefont
  {Rasing}},\ }\bibfield  {title} {\bibinfo {title} {Laser-induced
  magnetization dynamics and reversal in ferrimagnetic alloys},\ }\href
  {https://doi.org/10.1088/0034-4885/76/2/026501} {\bibfield  {journal}
  {\bibinfo  {journal} {Rep. Prog. Phys}\ }\textbf {\bibinfo {volume} {76}},\
  \bibinfo {pages} {026501} (\bibinfo {year} {2013})}\BibitemShut {NoStop}%
\bibitem [{\citenamefont {Walowski}\ and\ \citenamefont
  {Münzenberg}(2016)}]{walowski2016perspective}%
  \BibitemOpen
  \bibfield  {author} {\bibinfo {author} {\bibfnamefont {J.}~\bibnamefont
  {Walowski}}\ and\ \bibinfo {author} {\bibfnamefont {M.}~\bibnamefont
  {Münzenberg}},\ }\bibfield  {title} {\bibinfo {title} {{Perspective:
  Ultrafast magnetism and {THz} spintronics}},\ }\href
  {https://doi.org/10.1063/1.4958846} {\bibfield  {journal} {\bibinfo
  {journal} {J. Appl. Phys.}\ }\textbf {\bibinfo {volume} {120}},\ \bibinfo
  {pages} {140901} (\bibinfo {year} {2016})}\BibitemShut {NoStop}%
\bibitem [{\citenamefont {Buzzi}\ \emph {et~al.}(2018)\citenamefont {Buzzi},
  \citenamefont {F{\"o}rst}, \citenamefont {Mankowsky},\ and\ \citenamefont
  {Cavalleri}}]{buzzi2018probing}%
  \BibitemOpen
  \bibfield  {author} {\bibinfo {author} {\bibfnamefont {M.}~\bibnamefont
  {Buzzi}}, \bibinfo {author} {\bibfnamefont {M.}~\bibnamefont {F{\"o}rst}},
  \bibinfo {author} {\bibfnamefont {R.}~\bibnamefont {Mankowsky}},\ and\
  \bibinfo {author} {\bibfnamefont {A.}~\bibnamefont {Cavalleri}},\ }\bibfield
  {title} {\bibinfo {title} {Probing dynamics in quantum materials with
  femtosecond x-rays},\ }\href {https://doi.org/10.1038/s41578-018-0024-9}
  {\bibfield  {journal} {\bibinfo  {journal} {Nat. Rev. Mater.}\ }\textbf
  {\bibinfo {volume} {3}},\ \bibinfo {pages} {299} (\bibinfo {year}
  {2018})}\BibitemShut {NoStop}%
\bibitem [{\citenamefont {Carva}\ \emph {et~al.}(2017)\citenamefont {Carva},
  \citenamefont {Baláž},\ and\ \citenamefont {Radu}}]{CARVA2017291}%
  \BibitemOpen
  \bibfield  {author} {\bibinfo {author} {\bibfnamefont {K.}~\bibnamefont
  {Carva}}, \bibinfo {author} {\bibfnamefont {P.}~\bibnamefont {Baláž}},\
  and\ \bibinfo {author} {\bibfnamefont {I.}~\bibnamefont {Radu}},\ }\bibfield
  {title} {\bibinfo {title} {Chapter 2 - {L}aser-induced ultrafast magnetic
  phenomena}\ }(\bibinfo  {publisher} {Elsevier},\ \bibinfo {year} {2017})\
  pp.\ \bibinfo {pages} {291--463}\BibitemShut {NoStop}%
\bibitem [{\citenamefont {Wang}\ and\ \citenamefont
  {Liu}(2020)}]{wang2020ultrafast}%
  \BibitemOpen
  \bibfield  {author} {\bibinfo {author} {\bibfnamefont {C.}~\bibnamefont
  {Wang}}\ and\ \bibinfo {author} {\bibfnamefont {Y.}~\bibnamefont {Liu}},\
  }\bibfield  {title} {\bibinfo {title} {Ultrafast optical manipulation of
  magnetic order in ferromagnetic materials},\ }\href
  {https://doi.org/10.1186/s40580-020-00246-3} {\bibfield  {journal} {\bibinfo
  {journal} {Nano Converg.}\ }\textbf {\bibinfo {volume} {7}},\ \bibinfo
  {pages} {1} (\bibinfo {year} {2020})}\BibitemShut {NoStop}%
\bibitem [{\citenamefont {Scheid}\ \emph {et~al.}(2022)\citenamefont {Scheid},
  \citenamefont {Remy}, \citenamefont {Lebègue}, \citenamefont {Malinowski},\
  and\ \citenamefont {Mangin}}]{SCHEID2022169596}%
  \BibitemOpen
  \bibfield  {author} {\bibinfo {author} {\bibfnamefont {P.}~\bibnamefont
  {Scheid}}, \bibinfo {author} {\bibfnamefont {Q.}~\bibnamefont {Remy}},
  \bibinfo {author} {\bibfnamefont {S.}~\bibnamefont {Lebègue}}, \bibinfo
  {author} {\bibfnamefont {G.}~\bibnamefont {Malinowski}},\ and\ \bibinfo
  {author} {\bibfnamefont {S.}~\bibnamefont {Mangin}},\ }\bibfield  {title}
  {\bibinfo {title} {Light induced ultrafast magnetization dynamics in metallic
  compounds},\ }\href
  {https://doi.org/https://doi.org/10.1016/j.jmmm.2022.169596} {\bibfield
  {journal} {\bibinfo  {journal} {J. Magn. Magn. Mater.}\ }\textbf {\bibinfo
  {volume} {560}},\ \bibinfo {pages} {169596} (\bibinfo {year}
  {2022})}\BibitemShut {NoStop}%
\bibitem [{\citenamefont {Marti}\ \emph {et~al.}(2014)\citenamefont {Marti},
  \citenamefont {Fina}, \citenamefont {Frontera}, \citenamefont {Liu},
  \citenamefont {Wadley}, \citenamefont {He}, \citenamefont {Paull},
  \citenamefont {Clarkson}, \citenamefont {Kudrnovsk{\`y}}, \citenamefont
  {Turek}, \citenamefont {Kune{\v s}}, \citenamefont {Yi}, \citenamefont {Chu},
  \citenamefont {Nelson}, \citenamefont {You}, \citenamefont {Arenholz},
  \citenamefont {Salahuddin}, \citenamefont {Fontcuberta}, \citenamefont
  {Jungwirth},\ and\ \citenamefont {Ramesh}}]{marti2014room}%
  \BibitemOpen
  \bibfield  {author} {\bibinfo {author} {\bibfnamefont {X.}~\bibnamefont
  {Marti}}, \bibinfo {author} {\bibfnamefont {I.}~\bibnamefont {Fina}},
  \bibinfo {author} {\bibfnamefont {C.}~\bibnamefont {Frontera}}, \bibinfo
  {author} {\bibfnamefont {J.}~\bibnamefont {Liu}}, \bibinfo {author}
  {\bibfnamefont {P.}~\bibnamefont {Wadley}}, \bibinfo {author} {\bibfnamefont
  {Q.}~\bibnamefont {He}}, \bibinfo {author} {\bibfnamefont {R.~J.}\
  \bibnamefont {Paull}}, \bibinfo {author} {\bibfnamefont {J.~D.}\ \bibnamefont
  {Clarkson}}, \bibinfo {author} {\bibfnamefont {J.}~\bibnamefont
  {Kudrnovsk{\`y}}}, \bibinfo {author} {\bibfnamefont {I.}~\bibnamefont
  {Turek}}, \bibinfo {author} {\bibfnamefont {J.}~\bibnamefont {Kune{\v s}}},
  \bibinfo {author} {\bibfnamefont {D.}~\bibnamefont {Yi}}, \bibinfo {author}
  {\bibfnamefont {J.-H.}\ \bibnamefont {Chu}}, \bibinfo {author} {\bibfnamefont
  {C.~T.}\ \bibnamefont {Nelson}}, \bibinfo {author} {\bibfnamefont
  {L.}~\bibnamefont {You}}, \bibinfo {author} {\bibfnamefont {E.}~\bibnamefont
  {Arenholz}}, \bibinfo {author} {\bibfnamefont {S.}~\bibnamefont
  {Salahuddin}}, \bibinfo {author} {\bibfnamefont {J.}~\bibnamefont
  {Fontcuberta}}, \bibinfo {author} {\bibfnamefont {T.}~\bibnamefont
  {Jungwirth}},\ and\ \bibinfo {author} {\bibfnamefont {R.}~\bibnamefont
  {Ramesh}},\ }\bibfield  {title} {\bibinfo {title} {Room-temperature
  antiferromagnetic memory resistor},\ }\href
  {https://doi.org/10.1038/nmat3861} {\bibfield  {journal} {\bibinfo  {journal}
  {Nat. Mater.}\ }\textbf {\bibinfo {volume} {13}},\ \bibinfo {pages} {367}
  (\bibinfo {year} {2014})}\BibitemShut {NoStop}%
\bibitem [{\citenamefont {Wadley}\ \emph {et~al.}(2016)\citenamefont {Wadley},
  \citenamefont {Howells}, \citenamefont {Železný}, \citenamefont {Andrews},
  \citenamefont {Hills}, \citenamefont {Campion}, \citenamefont {Novák},
  \citenamefont {Olejník}, \citenamefont {Maccherozzi}, \citenamefont {Dhesi},
  \citenamefont {Martin}, \citenamefont {Wagner}, \citenamefont {Wunderlich},
  \citenamefont {Freimuth}, \citenamefont {Mokrousov}, \citenamefont {Kuneš},
  \citenamefont {Chauhan}, \citenamefont {Grzybowski}, \citenamefont
  {Rushforth}, \citenamefont {Edmonds}, \citenamefont {Gallagher},\ and\
  \citenamefont {Jungwirth}}]{wadley2016electrical}%
  \BibitemOpen
  \bibfield  {author} {\bibinfo {author} {\bibfnamefont {P.}~\bibnamefont
  {Wadley}}, \bibinfo {author} {\bibfnamefont {B.}~\bibnamefont {Howells}},
  \bibinfo {author} {\bibfnamefont {J.}~\bibnamefont {Železný}}, \bibinfo
  {author} {\bibfnamefont {C.}~\bibnamefont {Andrews}}, \bibinfo {author}
  {\bibfnamefont {V.}~\bibnamefont {Hills}}, \bibinfo {author} {\bibfnamefont
  {R.~P.}\ \bibnamefont {Campion}}, \bibinfo {author} {\bibfnamefont
  {V.}~\bibnamefont {Novák}}, \bibinfo {author} {\bibfnamefont
  {K.}~\bibnamefont {Olejník}}, \bibinfo {author} {\bibfnamefont
  {F.}~\bibnamefont {Maccherozzi}}, \bibinfo {author} {\bibfnamefont {S.~S.}\
  \bibnamefont {Dhesi}}, \bibinfo {author} {\bibfnamefont {S.~Y.}\ \bibnamefont
  {Martin}}, \bibinfo {author} {\bibfnamefont {T.}~\bibnamefont {Wagner}},
  \bibinfo {author} {\bibfnamefont {J.}~\bibnamefont {Wunderlich}}, \bibinfo
  {author} {\bibfnamefont {F.}~\bibnamefont {Freimuth}}, \bibinfo {author}
  {\bibfnamefont {Y.}~\bibnamefont {Mokrousov}}, \bibinfo {author}
  {\bibfnamefont {J.}~\bibnamefont {Kuneš}}, \bibinfo {author} {\bibfnamefont
  {J.~S.}\ \bibnamefont {Chauhan}}, \bibinfo {author} {\bibfnamefont {M.~J.}\
  \bibnamefont {Grzybowski}}, \bibinfo {author} {\bibfnamefont {A.~W.}\
  \bibnamefont {Rushforth}}, \bibinfo {author} {\bibfnamefont {K.~W.}\
  \bibnamefont {Edmonds}}, \bibinfo {author} {\bibfnamefont {B.~L.}\
  \bibnamefont {Gallagher}},\ and\ \bibinfo {author} {\bibfnamefont
  {T.}~\bibnamefont {Jungwirth}},\ }\bibfield  {title} {\bibinfo {title}
  {Electrical switching of an antiferromagnet},\ }\href
  {https://doi.org/10.1126/science.aab1031} {\bibfield  {journal} {\bibinfo
  {journal} {Science}\ }\textbf {\bibinfo {volume} {351}},\ \bibinfo {pages}
  {587} (\bibinfo {year} {2016})}\BibitemShut {NoStop}%
\bibitem [{\citenamefont {Kriegner}\ \emph {et~al.}(2016)\citenamefont
  {Kriegner}, \citenamefont {V{\'y}born{\'y}}, \citenamefont {Olejn{\'\i}k},
  \citenamefont {Reichlov{\'a}}, \citenamefont {Nov{\'a}k}, \citenamefont
  {Marti}, \citenamefont {Gazquez}, \citenamefont {Saidl}, \citenamefont
  {N{\v{e}}mec}, \citenamefont {Volobuev}, \citenamefont {Springholz},
  \citenamefont {Hol{\'y}},\ and\ \citenamefont
  {Jungwirth}}]{kriegner2016multiple}%
  \BibitemOpen
  \bibfield  {author} {\bibinfo {author} {\bibfnamefont {D.}~\bibnamefont
  {Kriegner}}, \bibinfo {author} {\bibfnamefont {K.}~\bibnamefont
  {V{\'y}born{\'y}}}, \bibinfo {author} {\bibfnamefont {K.}~\bibnamefont
  {Olejn{\'\i}k}}, \bibinfo {author} {\bibfnamefont {H.}~\bibnamefont
  {Reichlov{\'a}}}, \bibinfo {author} {\bibfnamefont {V.}~\bibnamefont
  {Nov{\'a}k}}, \bibinfo {author} {\bibfnamefont {X.}~\bibnamefont {Marti}},
  \bibinfo {author} {\bibfnamefont {J.}~\bibnamefont {Gazquez}}, \bibinfo
  {author} {\bibfnamefont {V.}~\bibnamefont {Saidl}}, \bibinfo {author}
  {\bibfnamefont {P.}~\bibnamefont {N{\v{e}}mec}}, \bibinfo {author}
  {\bibfnamefont {V.~V.}\ \bibnamefont {Volobuev}}, \bibinfo {author}
  {\bibfnamefont {G.}~\bibnamefont {Springholz}}, \bibinfo {author}
  {\bibfnamefont {V.}~\bibnamefont {Hol{\'y}}},\ and\ \bibinfo {author}
  {\bibfnamefont {T.}~\bibnamefont {Jungwirth}},\ }\bibfield  {title} {\bibinfo
  {title} {Multiple-stable anisotropic magnetoresistance memory in
  antiferromagnetic {M}n{T}e},\ }\href {https://doi.org/10.1038/ncomms11623}
  {\bibfield  {journal} {\bibinfo  {journal} {Nat. Commun.}\ }\textbf {\bibinfo
  {volume} {7}},\ \bibinfo {pages} {11623} (\bibinfo {year}
  {2016})}\BibitemShut {NoStop}%
\bibitem [{\citenamefont {Olejn{\'\i}k}\ \emph {et~al.}(2017)\citenamefont
  {Olejn{\'\i}k}, \citenamefont {Schuler}, \citenamefont {Mart{\'\i}},
  \citenamefont {Nov{\'a}k}, \citenamefont {Ka{\v{s}}par}, \citenamefont
  {Wadley}, \citenamefont {Campion}, \citenamefont {Edmonds}, \citenamefont
  {Gallagher}, \citenamefont {Garc{\'e}s}, \citenamefont {M}, \citenamefont
  {P},\ and\ \citenamefont {T}}]{olejnik2017antiferromagnetic}%
  \BibitemOpen
  \bibfield  {author} {\bibinfo {author} {\bibfnamefont {K.}~\bibnamefont
  {Olejn{\'\i}k}}, \bibinfo {author} {\bibfnamefont {V.}~\bibnamefont
  {Schuler}}, \bibinfo {author} {\bibfnamefont {X.}~\bibnamefont {Mart{\'\i}}},
  \bibinfo {author} {\bibfnamefont {V.}~\bibnamefont {Nov{\'a}k}}, \bibinfo
  {author} {\bibfnamefont {Z.}~\bibnamefont {Ka{\v{s}}par}}, \bibinfo {author}
  {\bibfnamefont {P.}~\bibnamefont {Wadley}}, \bibinfo {author} {\bibfnamefont
  {R.~P.}\ \bibnamefont {Campion}}, \bibinfo {author} {\bibfnamefont {K.~W.}\
  \bibnamefont {Edmonds}}, \bibinfo {author} {\bibfnamefont {B.~L.}\
  \bibnamefont {Gallagher}}, \bibinfo {author} {\bibfnamefont {J.}~\bibnamefont
  {Garc{\'e}s}}, \bibinfo {author} {\bibfnamefont {B.}~\bibnamefont {M}},
  \bibinfo {author} {\bibfnamefont {G.}~\bibnamefont {P}},\ and\ \bibinfo
  {author} {\bibfnamefont {J.}~\bibnamefont {T}},\ }\bibfield  {title}
  {\bibinfo {title} {Antiferromagnetic {C}u{M}n{A}s multi-level memory cell
  with microelectronic compatibility},\ }\href
  {https://doi.org/10.1038/ncomms15434} {\bibfield  {journal} {\bibinfo
  {journal} {Nat. Commun.}\ }\textbf {\bibinfo {volume} {8}},\ \bibinfo {pages}
  {15434} (\bibinfo {year} {2017})}\BibitemShut {NoStop}%
\bibitem [{\citenamefont {Jungwirth}\ \emph {et~al.}(2016)\citenamefont
  {Jungwirth}, \citenamefont {Marti}, \citenamefont {Wadley},\ and\
  \citenamefont {Wunderlich}}]{jungwirth2016antiferromagnetic}%
  \BibitemOpen
  \bibfield  {author} {\bibinfo {author} {\bibfnamefont {T.}~\bibnamefont
  {Jungwirth}}, \bibinfo {author} {\bibfnamefont {X.}~\bibnamefont {Marti}},
  \bibinfo {author} {\bibfnamefont {P.}~\bibnamefont {Wadley}},\ and\ \bibinfo
  {author} {\bibfnamefont {J.}~\bibnamefont {Wunderlich}},\ }\bibfield  {title}
  {\bibinfo {title} {Antiferromagnetic spintronics},\ }\href
  {https://doi.org/10.1038/nnano.2016.18} {\bibfield  {journal} {\bibinfo
  {journal} {Nat. Nano.}\ }\textbf {\bibinfo {volume} {11}},\ \bibinfo {pages}
  {231} (\bibinfo {year} {2016})}\BibitemShut {NoStop}%
\bibitem [{\citenamefont {Baltz}\ \emph {et~al.}(2018)\citenamefont {Baltz},
  \citenamefont {Manchon}, \citenamefont {Tsoi}, \citenamefont {Moriyama},
  \citenamefont {Ono},\ and\ \citenamefont
  {Tserkovnyak}}]{baltz2018antiferromagnetic}%
  \BibitemOpen
  \bibfield  {author} {\bibinfo {author} {\bibfnamefont {V.}~\bibnamefont
  {Baltz}}, \bibinfo {author} {\bibfnamefont {A.}~\bibnamefont {Manchon}},
  \bibinfo {author} {\bibfnamefont {M.}~\bibnamefont {Tsoi}}, \bibinfo {author}
  {\bibfnamefont {T.}~\bibnamefont {Moriyama}}, \bibinfo {author}
  {\bibfnamefont {T.}~\bibnamefont {Ono}},\ and\ \bibinfo {author}
  {\bibfnamefont {Y.}~\bibnamefont {Tserkovnyak}},\ }\bibfield  {title}
  {\bibinfo {title} {Antiferromagnetic spintronics},\ }\href
  {https://doi.org/10.1103/RevModPhys.90.015005} {\bibfield  {journal}
  {\bibinfo  {journal} {Rev. Mod. Phys.}\ }\textbf {\bibinfo {volume} {90}},\
  \bibinfo {pages} {015005} (\bibinfo {year} {2018})}\BibitemShut {NoStop}%
\bibitem [{\citenamefont {N{\v{e}}mec}\ \emph {et~al.}(2018)\citenamefont
  {N{\v{e}}mec}, \citenamefont {Fiebig}, \citenamefont {Kampfrath},\ and\
  \citenamefont {Kimel}}]{nvemec2018antiferromagnetic}%
  \BibitemOpen
  \bibfield  {author} {\bibinfo {author} {\bibfnamefont {P.}~\bibnamefont
  {N{\v{e}}mec}}, \bibinfo {author} {\bibfnamefont {M.}~\bibnamefont {Fiebig}},
  \bibinfo {author} {\bibfnamefont {T.}~\bibnamefont {Kampfrath}},\ and\
  \bibinfo {author} {\bibfnamefont {A.~V.}\ \bibnamefont {Kimel}},\ }\bibfield
  {title} {\bibinfo {title} {Antiferromagnetic opto-spintronics},\ }\href
  {https://doi.org/10.1038/s41567-018-0051-x} {\bibfield  {journal} {\bibinfo
  {journal} {Nat. Phys.}\ }\textbf {\bibinfo {volume} {14}},\ \bibinfo {pages}
  {229} (\bibinfo {year} {2018})}\BibitemShut {NoStop}%
\bibitem [{\citenamefont {Xiong}\ \emph {et~al.}(2022)\citenamefont {Xiong},
  \citenamefont {Jiang}, \citenamefont {Shi}, \citenamefont {Du}, \citenamefont
  {Yao}, \citenamefont {Guo}, \citenamefont {Zhu}, \citenamefont {Cao},
  \citenamefont {Peng}, \citenamefont {Cai}, \citenamefont {Zhu},\ and\
  \citenamefont {Zhao}}]{xiong2022antiferromagnetic}%
  \BibitemOpen
  \bibfield  {author} {\bibinfo {author} {\bibfnamefont {D.}~\bibnamefont
  {Xiong}}, \bibinfo {author} {\bibfnamefont {Y.}~\bibnamefont {Jiang}},
  \bibinfo {author} {\bibfnamefont {K.}~\bibnamefont {Shi}}, \bibinfo {author}
  {\bibfnamefont {A.}~\bibnamefont {Du}}, \bibinfo {author} {\bibfnamefont
  {Y.}~\bibnamefont {Yao}}, \bibinfo {author} {\bibfnamefont {Z.}~\bibnamefont
  {Guo}}, \bibinfo {author} {\bibfnamefont {D.}~\bibnamefont {Zhu}}, \bibinfo
  {author} {\bibfnamefont {K.}~\bibnamefont {Cao}}, \bibinfo {author}
  {\bibfnamefont {S.}~\bibnamefont {Peng}}, \bibinfo {author} {\bibfnamefont
  {W.}~\bibnamefont {Cai}}, \bibinfo {author} {\bibfnamefont {D.}~\bibnamefont
  {Zhu}},\ and\ \bibinfo {author} {\bibfnamefont {W.}~\bibnamefont {Zhao}},\
  }\bibfield  {title} {\bibinfo {title} {Antiferromagnetic spintronics: An
  overview and outlook},\ }\href
  {https://doi.org/https://doi.org/10.1016/j.fmre.2022.03.016} {\bibfield
  {journal} {\bibinfo  {journal} {Fundam. Res.}\ }\textbf {\bibinfo {volume}
  {2}},\ \bibinfo {pages} {522} (\bibinfo {year} {2022})}\BibitemShut {NoStop}%
\bibitem [{\citenamefont {Rettig}\ \emph {et~al.}(2016)\citenamefont {Rettig},
  \citenamefont {Dornes}, \citenamefont {Thielemann-K\"uhn}, \citenamefont
  {Pontius}, \citenamefont {Zabel}, \citenamefont {Schlagel}, \citenamefont
  {Lograsso}, \citenamefont {Chollet}, \citenamefont {Robert}, \citenamefont
  {Sikorski}, \citenamefont {Song}, \citenamefont {Glownia}, \citenamefont
  {Sch\"u\ss{}ler-Langeheine}, \citenamefont {Johnson},\ and\ \citenamefont
  {Staub}}]{rettig2016itinerant}%
  \BibitemOpen
  \bibfield  {author} {\bibinfo {author} {\bibfnamefont {L.}~\bibnamefont
  {Rettig}}, \bibinfo {author} {\bibfnamefont {C.}~\bibnamefont {Dornes}},
  \bibinfo {author} {\bibfnamefont {N.}~\bibnamefont {Thielemann-K\"uhn}},
  \bibinfo {author} {\bibfnamefont {N.}~\bibnamefont {Pontius}}, \bibinfo
  {author} {\bibfnamefont {H.}~\bibnamefont {Zabel}}, \bibinfo {author}
  {\bibfnamefont {D.~L.}\ \bibnamefont {Schlagel}}, \bibinfo {author}
  {\bibfnamefont {T.~A.}\ \bibnamefont {Lograsso}}, \bibinfo {author}
  {\bibfnamefont {M.}~\bibnamefont {Chollet}}, \bibinfo {author} {\bibfnamefont
  {A.}~\bibnamefont {Robert}}, \bibinfo {author} {\bibfnamefont
  {M.}~\bibnamefont {Sikorski}}, \bibinfo {author} {\bibfnamefont
  {S.}~\bibnamefont {Song}}, \bibinfo {author} {\bibfnamefont {J.~M.}\
  \bibnamefont {Glownia}}, \bibinfo {author} {\bibfnamefont {C.}~\bibnamefont
  {Sch\"u\ss{}ler-Langeheine}}, \bibinfo {author} {\bibfnamefont {S.~L.}\
  \bibnamefont {Johnson}},\ and\ \bibinfo {author} {\bibfnamefont
  {U.}~\bibnamefont {Staub}},\ }\bibfield  {title} {\bibinfo {title} {Itinerant
  and localized magnetization dynamics in antiferromagnetic {H}o},\ }\href
  {https://doi.org/10.1103/PhysRevLett.116.257202} {\bibfield  {journal}
  {\bibinfo  {journal} {Phys. Rev. Lett.}\ }\textbf {\bibinfo {volume} {116}},\
  \bibinfo {pages} {257202} (\bibinfo {year} {2016})}\BibitemShut {NoStop}%
\bibitem [{\citenamefont {Thielemann-K\"uhn}\ \emph {et~al.}(2017)\citenamefont
  {Thielemann-K\"uhn}, \citenamefont {Schick}, \citenamefont {Pontius},
  \citenamefont {Trabant}, \citenamefont {Mitzner}, \citenamefont {Holldack},
  \citenamefont {Zabel}, \citenamefont {F\"ohlisch},\ and\ \citenamefont
  {Sch\"u\ss{}ler-Langeheine}}]{thielemann2017ultrafast}%
  \BibitemOpen
  \bibfield  {author} {\bibinfo {author} {\bibfnamefont {N.}~\bibnamefont
  {Thielemann-K\"uhn}}, \bibinfo {author} {\bibfnamefont {D.}~\bibnamefont
  {Schick}}, \bibinfo {author} {\bibfnamefont {N.}~\bibnamefont {Pontius}},
  \bibinfo {author} {\bibfnamefont {C.}~\bibnamefont {Trabant}}, \bibinfo
  {author} {\bibfnamefont {R.}~\bibnamefont {Mitzner}}, \bibinfo {author}
  {\bibfnamefont {K.}~\bibnamefont {Holldack}}, \bibinfo {author}
  {\bibfnamefont {H.}~\bibnamefont {Zabel}}, \bibinfo {author} {\bibfnamefont
  {A.}~\bibnamefont {F\"ohlisch}},\ and\ \bibinfo {author} {\bibfnamefont
  {C.}~\bibnamefont {Sch\"u\ss{}ler-Langeheine}},\ }\bibfield  {title}
  {\bibinfo {title} {Ultrafast and energy-efficient quenching of spin order:
  Antiferromagnetism beats ferromagnetism},\ }\href
  {https://doi.org/10.1103/PhysRevLett.119.197202} {\bibfield  {journal}
  {\bibinfo  {journal} {Phys. Rev. Lett.}\ }\textbf {\bibinfo {volume} {119}},\
  \bibinfo {pages} {197202} (\bibinfo {year} {2017})}\BibitemShut {NoStop}%
\bibitem [{\citenamefont {Kumberg}\ \emph {et~al.}(2020)\citenamefont
  {Kumberg}, \citenamefont {Golias}, \citenamefont {Pontius}, \citenamefont
  {Hosseinifar}, \citenamefont {Frischmuth}, \citenamefont {Gelen},
  \citenamefont {Shinwari}, \citenamefont {Thakur}, \citenamefont
  {Sch\"u\ss{}ler-Langeheine}, \citenamefont {Oppeneer},\ and\ \citenamefont
  {Kuch}}]{kumberg2020accelerating}%
  \BibitemOpen
  \bibfield  {author} {\bibinfo {author} {\bibfnamefont {I.}~\bibnamefont
  {Kumberg}}, \bibinfo {author} {\bibfnamefont {E.}~\bibnamefont {Golias}},
  \bibinfo {author} {\bibfnamefont {N.}~\bibnamefont {Pontius}}, \bibinfo
  {author} {\bibfnamefont {R.}~\bibnamefont {Hosseinifar}}, \bibinfo {author}
  {\bibfnamefont {K.}~\bibnamefont {Frischmuth}}, \bibinfo {author}
  {\bibfnamefont {I.}~\bibnamefont {Gelen}}, \bibinfo {author} {\bibfnamefont
  {T.}~\bibnamefont {Shinwari}}, \bibinfo {author} {\bibfnamefont
  {S.}~\bibnamefont {Thakur}}, \bibinfo {author} {\bibfnamefont
  {C.}~\bibnamefont {Sch\"u\ss{}ler-Langeheine}}, \bibinfo {author}
  {\bibfnamefont {P.~M.}\ \bibnamefont {Oppeneer}},\ and\ \bibinfo {author}
  {\bibfnamefont {W.}~\bibnamefont {Kuch}},\ }\bibfield  {title} {\bibinfo
  {title} {Accelerating the laser-induced demagnetization of a ferromagnetic
  film by antiferromagnetic order in an adjacent layer},\ }\href
  {https://doi.org/10.1103/PhysRevB.102.214418} {\bibfield  {journal} {\bibinfo
   {journal} {Phys. Rev. B}\ }\textbf {\bibinfo {volume} {102}},\ \bibinfo
  {pages} {214418} (\bibinfo {year} {2020})}\BibitemShut {NoStop}%
\bibitem [{\citenamefont {Golias}\ \emph {et~al.}(2021)\citenamefont {Golias},
  \citenamefont {Kumberg}, \citenamefont {Gelen}, \citenamefont {Thakur},
  \citenamefont {G\"ordes}, \citenamefont {Hosseinifar}, \citenamefont
  {Guillet}, \citenamefont {Dewhurst}, \citenamefont {Sharma}, \citenamefont
  {Sch\"u\ss{}ler-Langeheine}, \citenamefont {Pontius},\ and\ \citenamefont
  {Kuch}}]{1854}%
  \BibitemOpen
  \bibfield  {author} {\bibinfo {author} {\bibfnamefont {E.}~\bibnamefont
  {Golias}}, \bibinfo {author} {\bibfnamefont {I.}~\bibnamefont {Kumberg}},
  \bibinfo {author} {\bibfnamefont {I.}~\bibnamefont {Gelen}}, \bibinfo
  {author} {\bibfnamefont {S.}~\bibnamefont {Thakur}}, \bibinfo {author}
  {\bibfnamefont {J.}~\bibnamefont {G\"ordes}}, \bibinfo {author}
  {\bibfnamefont {R.}~\bibnamefont {Hosseinifar}}, \bibinfo {author}
  {\bibfnamefont {Q.}~\bibnamefont {Guillet}}, \bibinfo {author} {\bibfnamefont
  {J.~K.}\ \bibnamefont {Dewhurst}}, \bibinfo {author} {\bibfnamefont
  {S.}~\bibnamefont {Sharma}}, \bibinfo {author} {\bibfnamefont
  {C.}~\bibnamefont {Sch\"u\ss{}ler-Langeheine}}, \bibinfo {author}
  {\bibfnamefont {N.}~\bibnamefont {Pontius}},\ and\ \bibinfo {author}
  {\bibfnamefont {W.}~\bibnamefont {Kuch}},\ }\bibfield  {title} {\bibinfo
  {title} {Ultrafast optically induced ferromagnetic state in an elemental
  antiferromagnet},\ }\href {https://doi.org/10.1103/PhysRevLett.126.107202}
  {\bibfield  {journal} {\bibinfo  {journal} {Phys. Rev. Lett.}\ }\textbf
  {\bibinfo {volume} {126}},\ \bibinfo {pages} {107202} (\bibinfo {year}
  {2021})}\BibitemShut {NoStop}%
\bibitem [{\citenamefont {Eschenlohr}(2020)}]{eschenlohr2020spin}%
  \BibitemOpen
  \bibfield  {author} {\bibinfo {author} {\bibfnamefont {A.}~\bibnamefont
  {Eschenlohr}},\ }\bibfield  {title} {\bibinfo {title} {Spin dynamics at
  interfaces on femtosecond timescales},\ }\href
  {https://doi.org/10.1088/1361-648X/abb519} {\bibfield  {journal} {\bibinfo
  {journal} {J. Condens. Matter Phys.}\ }\textbf {\bibinfo {volume} {33}},\
  \bibinfo {pages} {013001} (\bibinfo {year} {2020})}\BibitemShut {NoStop}%
\bibitem [{\citenamefont {Hellman}\ \emph {et~al.}(2017)\citenamefont
  {Hellman}, \citenamefont {Hoffmann}, \citenamefont {Tserkovnyak},
  \citenamefont {Beach}, \citenamefont {Fullerton}, \citenamefont {Leighton},
  \citenamefont {MacDonald}, \citenamefont {Ralph}, \citenamefont {Arena},
  \citenamefont {D\"urr}, \citenamefont {Fischer}, \citenamefont {Grollier},
  \citenamefont {Heremans}, \citenamefont {Jungwirth}, \citenamefont {Kimel},
  \citenamefont {Koopmans}, \citenamefont {Krivorotov}, \citenamefont {May},
  \citenamefont {Petford-Long}, \citenamefont {Rondinelli}, \citenamefont
  {Samarth}, \citenamefont {Schuller}, \citenamefont {Slavin}, \citenamefont
  {Stiles}, \citenamefont {Tchernyshyov}, \citenamefont {Thiaville},\ and\
  \citenamefont {Zink}}]{RevModPhys.89.025006}%
  \BibitemOpen
  \bibfield  {author} {\bibinfo {author} {\bibfnamefont {F.}~\bibnamefont
  {Hellman}}, \bibinfo {author} {\bibfnamefont {A.}~\bibnamefont {Hoffmann}},
  \bibinfo {author} {\bibfnamefont {Y.}~\bibnamefont {Tserkovnyak}}, \bibinfo
  {author} {\bibfnamefont {G.~S.~D.}\ \bibnamefont {Beach}}, \bibinfo {author}
  {\bibfnamefont {E.~E.}\ \bibnamefont {Fullerton}}, \bibinfo {author}
  {\bibfnamefont {C.}~\bibnamefont {Leighton}}, \bibinfo {author}
  {\bibfnamefont {A.~H.}\ \bibnamefont {MacDonald}}, \bibinfo {author}
  {\bibfnamefont {D.~C.}\ \bibnamefont {Ralph}}, \bibinfo {author}
  {\bibfnamefont {D.~A.}\ \bibnamefont {Arena}}, \bibinfo {author}
  {\bibfnamefont {H.~A.}\ \bibnamefont {D\"urr}}, \bibinfo {author}
  {\bibfnamefont {P.}~\bibnamefont {Fischer}}, \bibinfo {author} {\bibfnamefont
  {J.}~\bibnamefont {Grollier}}, \bibinfo {author} {\bibfnamefont {J.~P.}\
  \bibnamefont {Heremans}}, \bibinfo {author} {\bibfnamefont {T.}~\bibnamefont
  {Jungwirth}}, \bibinfo {author} {\bibfnamefont {A.~V.}\ \bibnamefont
  {Kimel}}, \bibinfo {author} {\bibfnamefont {B.}~\bibnamefont {Koopmans}},
  \bibinfo {author} {\bibfnamefont {I.~N.}\ \bibnamefont {Krivorotov}},
  \bibinfo {author} {\bibfnamefont {S.~J.}\ \bibnamefont {May}}, \bibinfo
  {author} {\bibfnamefont {A.~K.}\ \bibnamefont {Petford-Long}}, \bibinfo
  {author} {\bibfnamefont {J.~M.}\ \bibnamefont {Rondinelli}}, \bibinfo
  {author} {\bibfnamefont {N.}~\bibnamefont {Samarth}}, \bibinfo {author}
  {\bibfnamefont {I.~K.}\ \bibnamefont {Schuller}}, \bibinfo {author}
  {\bibfnamefont {A.~N.}\ \bibnamefont {Slavin}}, \bibinfo {author}
  {\bibfnamefont {M.~D.}\ \bibnamefont {Stiles}}, \bibinfo {author}
  {\bibfnamefont {O.}~\bibnamefont {Tchernyshyov}}, \bibinfo {author}
  {\bibfnamefont {A.}~\bibnamefont {Thiaville}},\ and\ \bibinfo {author}
  {\bibfnamefont {B.~L.}\ \bibnamefont {Zink}},\ }\bibfield  {title} {\bibinfo
  {title} {Interface-induced phenomena in magnetism},\ }\href
  {https://doi.org/10.1103/RevModPhys.89.025006} {\bibfield  {journal}
  {\bibinfo  {journal} {Rev. Mod. Phys.}\ }\textbf {\bibinfo {volume} {89}},\
  \bibinfo {pages} {025006} (\bibinfo {year} {2017})}\BibitemShut {NoStop}%
\bibitem [{\citenamefont {Saidl}\ \emph {et~al.}(2017)\citenamefont {Saidl},
  \citenamefont {N{\v{e}}mec}, \citenamefont {Wadley}, \citenamefont {Hills},
  \citenamefont {Campion}, \citenamefont {Nov{\'a}k}, \citenamefont {Edmonds},
  \citenamefont {Maccherozzi}, \citenamefont {Dhesi}, \citenamefont {Gallagher}
  \emph {et~al.}}]{saidl2017optical}%
  \BibitemOpen
  \bibfield  {author} {\bibinfo {author} {\bibfnamefont {V.}~\bibnamefont
  {Saidl}}, \bibinfo {author} {\bibfnamefont {P.}~\bibnamefont {N{\v{e}}mec}},
  \bibinfo {author} {\bibfnamefont {P.}~\bibnamefont {Wadley}}, \bibinfo
  {author} {\bibfnamefont {V.}~\bibnamefont {Hills}}, \bibinfo {author}
  {\bibfnamefont {R.}~\bibnamefont {Campion}}, \bibinfo {author} {\bibfnamefont
  {V.}~\bibnamefont {Nov{\'a}k}}, \bibinfo {author} {\bibfnamefont
  {K.}~\bibnamefont {Edmonds}}, \bibinfo {author} {\bibfnamefont
  {F.}~\bibnamefont {Maccherozzi}}, \bibinfo {author} {\bibfnamefont
  {S.}~\bibnamefont {Dhesi}}, \bibinfo {author} {\bibfnamefont
  {B.}~\bibnamefont {Gallagher}}, \emph {et~al.},\ }\bibfield  {title}
  {\bibinfo {title} {Optical determination of the {N}{\'e}el vector in a
  {C}u{M}n{A}s thin-film antiferromagnet},\ }\href
  {https://doi.org/10.1038/nphoton.2016.255} {\bibfield  {journal} {\bibinfo
  {journal} {Nat. Photonics}\ }\textbf {\bibinfo {volume} {11}},\ \bibinfo
  {pages} {91} (\bibinfo {year} {2017})}\BibitemShut {NoStop}%
\bibitem [{\citenamefont {Zheng}\ \emph {et~al.}(2018)\citenamefont {Zheng},
  \citenamefont {Shi}, \citenamefont {Li}, \citenamefont {Gu}, \citenamefont
  {Xia}, \citenamefont {Shen}, \citenamefont {Jin}, \citenamefont {Yuan},
  \citenamefont {Wu}, \citenamefont {Chen},\ and\ \citenamefont
  {Zhao}}]{zheng2018magneto}%
  \BibitemOpen
  \bibfield  {author} {\bibinfo {author} {\bibfnamefont {Z.}~\bibnamefont
  {Zheng}}, \bibinfo {author} {\bibfnamefont {J.~Y.}\ \bibnamefont {Shi}},
  \bibinfo {author} {\bibfnamefont {Q.}~\bibnamefont {Li}}, \bibinfo {author}
  {\bibfnamefont {T.}~\bibnamefont {Gu}}, \bibinfo {author} {\bibfnamefont
  {H.}~\bibnamefont {Xia}}, \bibinfo {author} {\bibfnamefont {L.~Q.}\
  \bibnamefont {Shen}}, \bibinfo {author} {\bibfnamefont {F.}~\bibnamefont
  {Jin}}, \bibinfo {author} {\bibfnamefont {H.~C.}\ \bibnamefont {Yuan}},
  \bibinfo {author} {\bibfnamefont {Y.~Z.}\ \bibnamefont {Wu}}, \bibinfo
  {author} {\bibfnamefont {L.~Y.}\ \bibnamefont {Chen}},\ and\ \bibinfo
  {author} {\bibfnamefont {H.~B.}\ \bibnamefont {Zhao}},\ }\bibfield  {title}
  {\bibinfo {title} {Magneto-optical probe of ultrafast spin dynamics in
  antiferromagnetic {C}o{O} thin films},\ }\href
  {https://doi.org/10.1103/PhysRevB.98.134409} {\bibfield  {journal} {\bibinfo
  {journal} {Phys. Rev. B}\ }\textbf {\bibinfo {volume} {98}},\ \bibinfo
  {pages} {134409} (\bibinfo {year} {2018})}\BibitemShut {NoStop}%
\bibitem [{\citenamefont {Oppeneer}\ \emph {et~al.}(2003)\citenamefont
  {Oppeneer}, \citenamefont {Mertins}, \citenamefont {Abramsohn}, \citenamefont
  {Gaupp}, \citenamefont {Gudat}, \citenamefont {Kune\ifmmode~\check{s}\else
  \v{s}\fi{}},\ and\ \citenamefont {Schneider}}]{Oppeneer2003}%
  \BibitemOpen
  \bibfield  {author} {\bibinfo {author} {\bibfnamefont {P.~M.}\ \bibnamefont
  {Oppeneer}}, \bibinfo {author} {\bibfnamefont {H.-C.}\ \bibnamefont
  {Mertins}}, \bibinfo {author} {\bibfnamefont {D.}~\bibnamefont {Abramsohn}},
  \bibinfo {author} {\bibfnamefont {A.}~\bibnamefont {Gaupp}}, \bibinfo
  {author} {\bibfnamefont {W.}~\bibnamefont {Gudat}}, \bibinfo {author}
  {\bibfnamefont {J.}~\bibnamefont {Kune\ifmmode~\check{s}\else \v{s}\fi{}}},\
  and\ \bibinfo {author} {\bibfnamefont {C.~M.}\ \bibnamefont {Schneider}},\
  }\bibfield  {title} {\bibinfo {title} {Buried antiferromagnetic films
  investigated by x-ray magneto-optical reflection spectroscopy},\ }\href
  {https://doi.org/10.1103/PhysRevB.67.052401} {\bibfield  {journal} {\bibinfo
  {journal} {Phys. Rev. B}\ }\textbf {\bibinfo {volume} {67}},\ \bibinfo
  {pages} {052401} (\bibinfo {year} {2003})}\BibitemShut {NoStop}%
\bibitem [{\citenamefont {Ma}\ \emph {et~al.}(2015)\citenamefont {Ma},
  \citenamefont {Fang}, \citenamefont {Li}, \citenamefont {Zhu}, \citenamefont
  {Yang}, \citenamefont {Wu}, \citenamefont {Zhao},\ and\ \citenamefont
  {L{\"u}pke}}]{ma2015ultrafast}%
  \BibitemOpen
  \bibfield  {author} {\bibinfo {author} {\bibfnamefont {X.}~\bibnamefont
  {Ma}}, \bibinfo {author} {\bibfnamefont {F.}~\bibnamefont {Fang}}, \bibinfo
  {author} {\bibfnamefont {Q.}~\bibnamefont {Li}}, \bibinfo {author}
  {\bibfnamefont {J.}~\bibnamefont {Zhu}}, \bibinfo {author} {\bibfnamefont
  {Y.}~\bibnamefont {Yang}}, \bibinfo {author} {\bibfnamefont {Y.~Z.}\
  \bibnamefont {Wu}}, \bibinfo {author} {\bibfnamefont {H.~B.}\ \bibnamefont
  {Zhao}},\ and\ \bibinfo {author} {\bibfnamefont {G.}~\bibnamefont
  {L{\"u}pke}},\ }\bibfield  {title} {\bibinfo {title} {Ultrafast spin
  exchange-coupling torque via photo-excited charge-transfer processes},\
  }\href {https://doi.org/10.1038/ncomms9800} {\bibfield  {journal} {\bibinfo
  {journal} {Nat. Commun.}\ }\textbf {\bibinfo {volume} {6}},\ \bibinfo {pages}
  {8800} (\bibinfo {year} {2015})}\BibitemShut {NoStop}%
\bibitem [{\citenamefont {Mertins}\ \emph {et~al.}(2002)\citenamefont
  {Mertins}, \citenamefont {Abramsohn}, \citenamefont {Gaupp}, \citenamefont
  {Sch\"afers}, \citenamefont {Gudat}, \citenamefont {Zaharko}, \citenamefont
  {Grimmer},\ and\ \citenamefont {Oppeneer}}]{Mertins2002}%
  \BibitemOpen
  \bibfield  {author} {\bibinfo {author} {\bibfnamefont {H.-C.}\ \bibnamefont
  {Mertins}}, \bibinfo {author} {\bibfnamefont {D.}~\bibnamefont {Abramsohn}},
  \bibinfo {author} {\bibfnamefont {A.}~\bibnamefont {Gaupp}}, \bibinfo
  {author} {\bibfnamefont {F.}~\bibnamefont {Sch\"afers}}, \bibinfo {author}
  {\bibfnamefont {W.}~\bibnamefont {Gudat}}, \bibinfo {author} {\bibfnamefont
  {O.}~\bibnamefont {Zaharko}}, \bibinfo {author} {\bibfnamefont
  {H.}~\bibnamefont {Grimmer}},\ and\ \bibinfo {author} {\bibfnamefont {P.~M.}\
  \bibnamefont {Oppeneer}},\ }\bibfield  {title} {\bibinfo {title} {Resonant
  magnetic reflection coefficients at the {F}e $2p$ edge obtained with linearly
  and circularly polarized soft x rays},\ }\href
  {https://doi.org/10.1103/PhysRevB.66.184404} {\bibfield  {journal} {\bibinfo
  {journal} {Phys. Rev. B}\ }\textbf {\bibinfo {volume} {66}},\ \bibinfo
  {pages} {184404} (\bibinfo {year} {2002})}\BibitemShut {NoStop}%
\bibitem [{\citenamefont {Abrudan}\ \emph {et~al.}(2008)\citenamefont
  {Abrudan}, \citenamefont {Miguel}, \citenamefont {Bernien}, \citenamefont
  {Tieg}, \citenamefont {Piantek}, \citenamefont {Kirschner},\ and\
  \citenamefont {Kuch}}]{abrudan2008structural}%
  \BibitemOpen
  \bibfield  {author} {\bibinfo {author} {\bibfnamefont {R.}~\bibnamefont
  {Abrudan}}, \bibinfo {author} {\bibfnamefont {J.}~\bibnamefont {Miguel}},
  \bibinfo {author} {\bibfnamefont {M.}~\bibnamefont {Bernien}}, \bibinfo
  {author} {\bibfnamefont {C.}~\bibnamefont {Tieg}}, \bibinfo {author}
  {\bibfnamefont {M.}~\bibnamefont {Piantek}}, \bibinfo {author} {\bibfnamefont
  {J.}~\bibnamefont {Kirschner}},\ and\ \bibinfo {author} {\bibfnamefont
  {W.}~\bibnamefont {Kuch}},\ }\bibfield  {title} {\bibinfo {title} {Structural
  and magnetic properties of epitaxial {F}e/{C}o{O} bilayers on {A}g(001)},\
  }\href {https://doi.org/10.1103/PhysRevB.77.014411} {\bibfield  {journal}
  {\bibinfo  {journal} {Phys. Rev. B}\ }\textbf {\bibinfo {volume} {77}},\
  \bibinfo {pages} {014411} (\bibinfo {year} {2008})}\BibitemShut {NoStop}%
\bibitem [{sup()}]{supp}%
  \BibitemOpen
  \href@noop {} {\bibinfo {title} {See {S}upplemental {M}aterial at [{URL} will
  be inserted by publisher] for details of sample preparation, beamline
  parameters, details and parameters of simulations, and additional
  simulations.}}\BibitemShut {Stop}%
\bibitem [{\citenamefont {Miguel}\ \emph {et~al.}(2009)\citenamefont {Miguel},
  \citenamefont {Abrudan}, \citenamefont {Bernien}, \citenamefont {Piantek},
  \citenamefont {Tieg}, \citenamefont {Kirschner},\ and\ \citenamefont
  {Kuch}}]{miguel2009magnetic}%
  \BibitemOpen
  \bibfield  {author} {\bibinfo {author} {\bibfnamefont {J.}~\bibnamefont
  {Miguel}}, \bibinfo {author} {\bibfnamefont {R.}~\bibnamefont {Abrudan}},
  \bibinfo {author} {\bibfnamefont {M.}~\bibnamefont {Bernien}}, \bibinfo
  {author} {\bibfnamefont {M.}~\bibnamefont {Piantek}}, \bibinfo {author}
  {\bibfnamefont {C.}~\bibnamefont {Tieg}}, \bibinfo {author} {\bibfnamefont
  {J.}~\bibnamefont {Kirschner}},\ and\ \bibinfo {author} {\bibfnamefont
  {W.}~\bibnamefont {Kuch}},\ }\bibfield  {title} {\bibinfo {title} {Magnetic
  domain coupling study in single-crystalline {F}e/{C}o{O} bilayers},\ }\href
  {https://doi.org/10.1088/0953-8984/21/18/185004} {\bibfield  {journal}
  {\bibinfo  {journal} {J. Condens. Matter Phys.}\ }\textbf {\bibinfo {volume}
  {21}},\ \bibinfo {pages} {185004} (\bibinfo {year} {2009})}\BibitemShut
  {NoStop}%
\bibitem [{\citenamefont {Wust}\ \emph {et~al.}(2022)\citenamefont {Wust},
  \citenamefont {Seibel}, \citenamefont {Meer}, \citenamefont {Herrgen},
  \citenamefont {Schmitt}, \citenamefont {Baldrati}, \citenamefont {Ramos},
  \citenamefont {Kikkawa}, \citenamefont {Saitoh}, \citenamefont {Gomonay},
  \citenamefont {Sinova}, \citenamefont {Mokrousov}, \citenamefont {Schneider},
  \citenamefont {Kläui}, \citenamefont {Rethfeld}, \citenamefont
  {Stadtmüller},\ and\ \citenamefont {Aeschlimann}}]{wust2022indirect}%
  \BibitemOpen
  \bibfield  {author} {\bibinfo {author} {\bibfnamefont {S.}~\bibnamefont
  {Wust}}, \bibinfo {author} {\bibfnamefont {C.}~\bibnamefont {Seibel}},
  \bibinfo {author} {\bibfnamefont {H.}~\bibnamefont {Meer}}, \bibinfo {author}
  {\bibfnamefont {P.}~\bibnamefont {Herrgen}}, \bibinfo {author} {\bibfnamefont
  {C.}~\bibnamefont {Schmitt}}, \bibinfo {author} {\bibfnamefont
  {L.}~\bibnamefont {Baldrati}}, \bibinfo {author} {\bibfnamefont
  {R.}~\bibnamefont {Ramos}}, \bibinfo {author} {\bibfnamefont
  {T.}~\bibnamefont {Kikkawa}}, \bibinfo {author} {\bibfnamefont
  {E.}~\bibnamefont {Saitoh}}, \bibinfo {author} {\bibfnamefont
  {O.}~\bibnamefont {Gomonay}}, \bibinfo {author} {\bibfnamefont
  {J.}~\bibnamefont {Sinova}}, \bibinfo {author} {\bibfnamefont
  {Y.}~\bibnamefont {Mokrousov}}, \bibinfo {author} {\bibfnamefont {H.~C.}\
  \bibnamefont {Schneider}}, \bibinfo {author} {\bibfnamefont {M.}~\bibnamefont
  {Kläui}}, \bibinfo {author} {\bibfnamefont {B.}~\bibnamefont {Rethfeld}},
  \bibinfo {author} {\bibfnamefont {B.}~\bibnamefont {Stadtmüller}},\ and\
  \bibinfo {author} {\bibfnamefont {M.}~\bibnamefont {Aeschlimann}},\
  }\href@noop {} {\bibinfo {title} {Indirect optical manipulation of the
  antiferromagnetic order of insulating {N}i{O} by ultrafast interfacial energy
  transfer}} (\bibinfo {year} {2022}),\ \Eprint
  {https://arxiv.org/abs/2205.02686} {arXiv:2205.02686 [cond-mat.mtrl-sci]}
  \BibitemShut {NoStop}%
\bibitem [{\citenamefont {Nowak}(2007)}]{Nowak2007}%
  \BibitemOpen
  \bibfield  {author} {\bibinfo {author} {\bibfnamefont {U.}~\bibnamefont
  {Nowak}},\ }\bibinfo {title} {Classical spin models},\ in\ \href
  {https://doi.org/https://doi.org/10.1002/9780470022184.hmm205} {\emph
  {\bibinfo {booktitle} {Handbook of Magnetism and Advanced Magnetic
  Materials}}},\ \bibinfo {editor} {edited by\ \bibinfo {editor} {\bibfnamefont
  {H.}~\bibnamefont {Kronm{\"u}ller}}\ and\ \bibinfo {editor} {\bibfnamefont
  {S.}~\bibnamefont {Parkin}}}\ (\bibinfo  {publisher} {John Wiley \& Sons,
  Ltd},\ \bibinfo {year} {2007})\BibitemShut {NoStop}%
\bibitem [{\citenamefont {Kazantseva}\ \emph {et~al.}(2007)\citenamefont
  {Kazantseva}, \citenamefont {Nowak}, \citenamefont {Chantrell}, \citenamefont
  {Hohlfeld},\ and\ \citenamefont {Rebei}}]{Kazantseva2008}%
  \BibitemOpen
  \bibfield  {author} {\bibinfo {author} {\bibfnamefont {N.}~\bibnamefont
  {Kazantseva}}, \bibinfo {author} {\bibfnamefont {U.}~\bibnamefont {Nowak}},
  \bibinfo {author} {\bibfnamefont {R.~W.}\ \bibnamefont {Chantrell}}, \bibinfo
  {author} {\bibfnamefont {J.}~\bibnamefont {Hohlfeld}},\ and\ \bibinfo
  {author} {\bibfnamefont {A.}~\bibnamefont {Rebei}},\ }\bibfield  {title}
  {\bibinfo {title} {Slow recovery of the magnetisation after a sub-picosecond
  heat pulse},\ }\href {https://doi.org/10.1209/0295-5075/81/27004} {\bibfield
  {journal} {\bibinfo  {journal} {Europhysics Letters}\ }\textbf {\bibinfo
  {volume} {81}},\ \bibinfo {pages} {27004} (\bibinfo {year}
  {2007})}\BibitemShut {NoStop}%
\bibitem [{\citenamefont {Garanin}(1996)}]{Garanin1996}%
  \BibitemOpen
  \bibfield  {author} {\bibinfo {author} {\bibfnamefont {D.~A.}\ \bibnamefont
  {Garanin}},\ }\bibfield  {title} {\bibinfo {title} {Self-consistent
  {G}aussian approximation for classical spin systems: Thermodynamics},\ }\href
  {https://doi.org/10.1103/PhysRevB.53.11593} {\bibfield  {journal} {\bibinfo
  {journal} {Phys. Rev. B}\ }\textbf {\bibinfo {volume} {53}},\ \bibinfo
  {pages} {11593} (\bibinfo {year} {1996})}\BibitemShut {NoStop}%
\bibitem [{\citenamefont {Landau}\ and\ \citenamefont
  {Lifshitz}(1935)}]{Landau1935}%
  \BibitemOpen
  \bibfield  {author} {\bibinfo {author} {\bibfnamefont {L.~D.}\ \bibnamefont
  {Landau}}\ and\ \bibinfo {author} {\bibfnamefont {E.~M.}\ \bibnamefont
  {Lifshitz}},\ }\bibfield  {title} {\bibinfo {title} {On the theory of the
  dispersion of magnetic permeability in ferromagnetic bodies},\ }\href@noop {}
  {\bibfield  {journal} {\bibinfo  {journal} {Phys. Z. Sowjetunion}\ }\textbf
  {\bibinfo {volume} {8}},\ \bibinfo {pages} {101} (\bibinfo {year}
  {1935})}\BibitemShut {NoStop}%
\bibitem [{\citenamefont {Brown}(1963)}]{brown1963thermal}%
  \BibitemOpen
  \bibfield  {author} {\bibinfo {author} {\bibfnamefont {W.~F.}\ \bibnamefont
  {Brown}},\ }\bibfield  {title} {\bibinfo {title} {Thermal fluctuations of a
  single-domain particle},\ }\href {https://doi.org/10.1103/PhysRev.130.1677}
  {\bibfield  {journal} {\bibinfo  {journal} {Phys. Rev.}\ }\textbf {\bibinfo
  {volume} {130}},\ \bibinfo {pages} {1677} (\bibinfo {year}
  {1963})}\BibitemShut {NoStop}%
\bibitem [{\citenamefont {Gilbert}(2004)}]{Gilbert2004}%
  \BibitemOpen
  \bibfield  {author} {\bibinfo {author} {\bibfnamefont {T.~L.}\ \bibnamefont
  {Gilbert}},\ }\bibfield  {title} {\bibinfo {title} {A phenomenological theory
  of damping in ferromagnetic materials},\ }\href
  {https://doi.org/10.1109/TMAG.2004.836740} {\bibfield  {journal} {\bibinfo
  {journal} {IEEE Trans. Magn.}\ }\textbf {\bibinfo {volume} {40}},\ \bibinfo
  {pages} {3443} (\bibinfo {year} {2004})}\BibitemShut {NoStop}%
\bibitem [{\citenamefont {Zahn}\ \emph {et~al.}(2021)\citenamefont {Zahn},
  \citenamefont {Jakobs}, \citenamefont {Windsor}, \citenamefont {Seiler},
  \citenamefont {Vasileiadis}, \citenamefont {Butcher}, \citenamefont {Qi},
  \citenamefont {Engel}, \citenamefont {Atxitia}, \citenamefont {Vorberger},\
  and\ \citenamefont {Ernstorfer}}]{Zahn2021}%
  \BibitemOpen
  \bibfield  {author} {\bibinfo {author} {\bibfnamefont {D.}~\bibnamefont
  {Zahn}}, \bibinfo {author} {\bibfnamefont {F.}~\bibnamefont {Jakobs}},
  \bibinfo {author} {\bibfnamefont {Y.~W.}\ \bibnamefont {Windsor}}, \bibinfo
  {author} {\bibfnamefont {H.}~\bibnamefont {Seiler}}, \bibinfo {author}
  {\bibfnamefont {T.}~\bibnamefont {Vasileiadis}}, \bibinfo {author}
  {\bibfnamefont {T.~A.}\ \bibnamefont {Butcher}}, \bibinfo {author}
  {\bibfnamefont {Y.}~\bibnamefont {Qi}}, \bibinfo {author} {\bibfnamefont
  {D.}~\bibnamefont {Engel}}, \bibinfo {author} {\bibfnamefont
  {U.}~\bibnamefont {Atxitia}}, \bibinfo {author} {\bibfnamefont
  {J.}~\bibnamefont {Vorberger}},\ and\ \bibinfo {author} {\bibfnamefont
  {R.}~\bibnamefont {Ernstorfer}},\ }\bibfield  {title} {\bibinfo {title}
  {Lattice dynamics and ultrafast energy flow between electrons, spins, and
  phonons in a 3d ferromagnet},\ }\href
  {https://doi.org/10.1103/PhysRevResearch.3.023032} {\bibfield  {journal}
  {\bibinfo  {journal} {Phys. Rev. Res.}\ }\textbf {\bibinfo {volume} {3}},\
  \bibinfo {pages} {023032} (\bibinfo {year} {2021})}\BibitemShut {NoStop}%
\bibitem [{\citenamefont {Zahn}\ \emph {et~al.}(2022)\citenamefont {Zahn},
  \citenamefont {Jakobs}, \citenamefont {Seiler}, \citenamefont {Butcher},
  \citenamefont {Engel}, \citenamefont {Vorberger}, \citenamefont {Atxitia},
  \citenamefont {Windsor},\ and\ \citenamefont {Ernstorfer}}]{Zahn2022}%
  \BibitemOpen
  \bibfield  {author} {\bibinfo {author} {\bibfnamefont {D.}~\bibnamefont
  {Zahn}}, \bibinfo {author} {\bibfnamefont {F.}~\bibnamefont {Jakobs}},
  \bibinfo {author} {\bibfnamefont {H.}~\bibnamefont {Seiler}}, \bibinfo
  {author} {\bibfnamefont {T.~A.}\ \bibnamefont {Butcher}}, \bibinfo {author}
  {\bibfnamefont {D.}~\bibnamefont {Engel}}, \bibinfo {author} {\bibfnamefont
  {J.}~\bibnamefont {Vorberger}}, \bibinfo {author} {\bibfnamefont
  {U.}~\bibnamefont {Atxitia}}, \bibinfo {author} {\bibfnamefont {Y.~W.}\
  \bibnamefont {Windsor}},\ and\ \bibinfo {author} {\bibfnamefont
  {R.}~\bibnamefont {Ernstorfer}},\ }\bibfield  {title} {\bibinfo {title}
  {Intrinsic energy flow in laser-excited $3d$ ferromagnets},\ }\href
  {https://doi.org/10.1103/PhysRevResearch.4.013104} {\bibfield  {journal}
  {\bibinfo  {journal} {Phys. Rev. Res.}\ }\textbf {\bibinfo {volume} {4}},\
  \bibinfo {pages} {013104} (\bibinfo {year} {2022})}\BibitemShut {NoStop}%
\bibitem [{\citenamefont {Chen}\ \emph {et~al.}(2015)\citenamefont {Chen},
  \citenamefont {Sigrist}, \citenamefont {Sinova},\ and\ \citenamefont
  {Manske}}]{Chen2015}%
  \BibitemOpen
  \bibfield  {author} {\bibinfo {author} {\bibfnamefont {W.}~\bibnamefont
  {Chen}}, \bibinfo {author} {\bibfnamefont {M.}~\bibnamefont {Sigrist}},
  \bibinfo {author} {\bibfnamefont {J.}~\bibnamefont {Sinova}},\ and\ \bibinfo
  {author} {\bibfnamefont {D.}~\bibnamefont {Manske}},\ }\bibfield  {title}
  {\bibinfo {title} {Minimal model of spin-transfer torque and spin pumping
  caused by the spin {H}all effect},\ }\href
  {https://doi.org/10.1103/PhysRevLett.115.217203} {\bibfield  {journal}
  {\bibinfo  {journal} {Phys. Rev. Lett.}\ }\textbf {\bibinfo {volume} {115}},\
  \bibinfo {pages} {217203} (\bibinfo {year} {2015})}\BibitemShut {NoStop}%
\end{thebibliography}

\providecommand{\noopsort}[1]{}\providecommand{\singleletter}[1]{#1}%

\end{document}


\title{Element-selective probing of ultrafast ferromagnetic--antiferromagnetic order dynamics in Fe/CoO bilayers
\\--- Supplemental Material ---}

\author{Chowdhury S. Awsaf}
\affiliation{Institut für Experimentalphysik, Freie Universität Berlin, Arnimallee 14, 14195 Berlin, Germany}

\author{Sangeeta Thakur}
\affiliation{Institut für Experimentalphysik, Freie Universität Berlin, Arnimallee 14, 14195 Berlin, Germany}

\author{Markus Weißenhofer}
\affiliation{Institut für Experimentalphysik, Freie Universität Berlin, Arnimallee 14, 14195 Berlin, Germany}
\affiliation{Department of Physics and Astronomy, Uppsala University, Box 516, 75120 Uppsala, Sweden}

\author{Jendrik Gördes}
\affiliation{Institut für Experimentalphysik, Freie Universität Berlin, Arnimallee 14, 14195 Berlin, Germany}

\author{Marcel Walter}
\affiliation{Institut für Experimentalphysik, Freie Universität Berlin, Arnimallee 14, 14195 Berlin, Germany}

\author{Niko Pontius}
\affiliation{Helmholtz-Zentrum Berlin für Materialien und Energie, Albert-Einstein-Straße 15, 12489 Berlin, Germany}

\author{Christian Schüßler-Langeheine}
\affiliation{Helmholtz-Zentrum Berlin für Materialien und Energie, Albert-Einstein-Straße 15, 12489 Berlin, Germany}

\author{Peter M. Oppeneer}
\affiliation{Department of Physics and Astronomy, Uppsala University, Box 516, 75120 Uppsala, Sweden}

\author{Wolfgang Kuch}
\affiliation{Institut für Experimentalphysik, Freie Universität Berlin, Arnimallee 14, 14195 Berlin, Germany}

\maketitle

\subsection*{Sample preparation}
The Ag(001) substrate was cleaned by Ar sputtering at an Ar pressure of 10$^{-5}$ mbar followed by annealing at 750 K for 30 minutes. Co was then thermally evaporated onto the substrate, using a commercial e-beam evaporator, in an O${_2}$ atmosphere of 10$^{-6}$ mbar at 450 K sample temperature to prevent the formation of clusters or islands with different crystallographic directions and to enhance Co oxidation at the surface \cite{abrudan2008structural}. After 30 minutes of post-annealing at around 500 K in the presence 10$^{-6}$ mbar of O${_2}$, the Fe film was deposited at room temperature by thermal deposition from another commercial e-beam evaporator. Low-energy electron diffraction (LEED) was used to check the surface integrity of each layer and Auger electron spectroscopy (AES) was used to verify the cleanliness and thickness of each layer.  The accuracy of the AES thickness calibration is about $\pm 0.5$ ML for CoO and $\pm 1$ ML for Fe.  After deposition, \textit{in-situ} longitudinal magneto-optical Kerr effect was used to characterize the temperature-dependent static magnetization of the bilayer along the Ag [110] direction, which corresponds to the [100] Fe magnetization easy axis.

\subsection*{Reflectivity measurement optimization}
To measure the antiferromagnetic order dynamics, an x-ray magnetic spectroscopy sensitive to $M^2$ is required, $M$ being the sublattice magnetization. This is for example given by the x-ray magnetic linear dichroism (XMLD) or x-ray Voigt effect measured in transmission or absorption \cite{Mertins2001,abrudan2008structural}. However, these magneto-x-ray spectroscopies are not well suited for buried antiferromagnets (AFM's) or AFM's on thick metallic substrates. For such samples the x-ray magnetic linear dichroism measured in reflection (R-XMLD) provides a very suitable technique \cite{Oppeneer2003}. This magneto-x-ray spectroscopy can be used to probe buried AFM layers, as is the case for our Fe/CoO sample, and it has the advantage that the magnetic contrast can be optimized by varying the angle of incidence. 

To determine the angle and photon energy to obtain large magnetic contrast during the measurements, a series of static reflectivity scans were recorded beforehand,  without compromising the incident photon count. The angle and energy values with high figures of merit (expressed as the product of the square of the static dichroic magnitude and the reflected photon intensity \cite{kumberg2023ultrafast}) were maximized to reduce the required data acquisition time.

A BESSY II acquisition software was used to calculate the incident laser fluence from the average laser power of 1 W, incident angle, and spot size. The numbers given in the manuscript correspond to the incident fluence in the center of the laser spot at the surface of the sample.  Due to the accuracy of the spot-size determination, there is a systematic error of up to 30\% in the fluence, which might be different at 400 and 800 nm pump wavelength.  The experimental time resolution was estimated to be 120 fs (60 fs pump-pulse and 100 fs probe-pulse width) with an energy resolution of E/\textDelta E $\simeq$ 250. A series of static reflectivity scans was recorded, beforehand, to determine the angle and photon energy to obtain large magnetic contrast during the measurements without compromising the incident photon count. The angle and energy values with high figures of merit (expressed as the product of the square of the static dichroic magnitude and the reflected photon intensity \cite{kumberg2023ultrafast}) were maximized to reduce the required data acquisition time.

\subsection*{Magnetic asymmetry determination}

The dynamic magnetic signals or magneto-x-ray asymmetries were determined from the difference between the reflected signals for the two Fe magnetization directions with and without laser excitation. 

For CoO, the magnetic R-XMLD asymmetry at the Co $L_2$ edge is given by
\begin{small}
\begin{equation}
        Asymmetry_\mathrm{CoO}(t) = \frac{R^\parallel_{pumped}(t) - R^\perp_{pumped}(t)}{R^\parallel_{unpumped}(t) - R^\perp_{unpumped}(t)} \, , 
        \label{CoO_asymmetry}
\end{equation}
\end{small}
where $R^{\parallel}$ and $R^{\perp}$ are  the reflected intensities, measured for magnetic field directions parallel and perpendicular to the electric field vector of the linearly polarized x rays, respectively.

Similarly for Fe, the R-XMCD asymmetry at the $L_3$ edge can be obtained as

\begin{small}
\begin{equation}
        Asymmetry_\mathrm{Fe}(t) = \frac{R^{\uparrow \uparrow}_{ pumped}(t) - R^{\uparrow \downarrow}_{pumped}(t)}{R^{\uparrow \uparrow}_{unpumped}(t) - R^{\uparrow \downarrow}_{unpumped}(t)} \, , 
        \label{Fe_asymmetry}
\end{equation}
\end{small}
where $R^{\uparrow \uparrow}$ and $R^{\uparrow \downarrow}$ are the reflectivities for magnetic field directions parallel and antiparallel to the helicity of the circularly polarized x rays, respectively.

\subsection*{Evaluation of demagnetization times and amplitudes}

To extract the (sublattice) demagnetization times from the experimental data, we fitted the time traces of the magnetic dichroism with a double-exponential function.  
The phenomenological fitting model is given by

\begin{small}
\begin{equation}
        M(t) = G(t)*\left(\Theta(t-t_0)\left[A_{de}(e^{-\frac{t-t_0}{\tau_{de}}}-1) - A_{re}(e^{-\frac{t-t_0}{\tau_{re}}}-1)+1\right]\right) ,
        \label{expeq}
\end{equation}
\end{small}
\noindent
where $\tau_{de}$ and $\tau_{re}$ are the de- and remagnetization times, respectively, $t_0$ the time when the laser pulse hits the sample, $A_{de}$ and $A_{re}$ are magnetic order reduction and recovery amplitudes, respectively, $G(t)$ is the Gaussian response function of width 120 fs and $\theta(t-t_0)$ is the Heaviside step function.  The $*$ symbol represents the convolution.

\begin{figure}[b]
\includegraphics[width=0.85\linewidth]{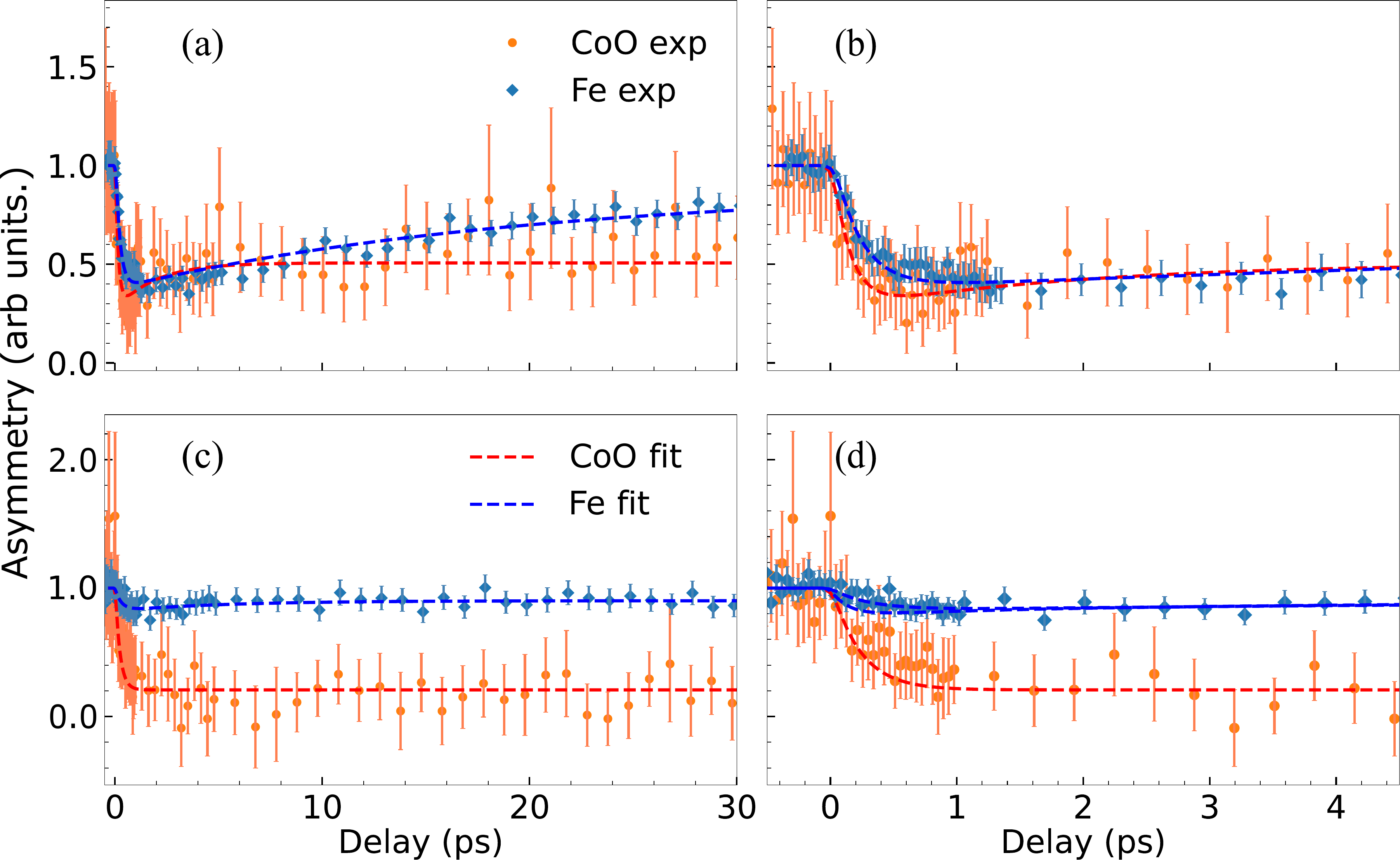}
\caption{\label{expfit} Experimental time traces of the reflection magnetic circular dichroism at the Fe $L_3$ edge and the magnetic linear dichroism at the Co $L_2$ edge with fits to double-exponential functions (dashed lines). Both short-range (a and c) and long-range (b and d) delay-time graphs illustrate the different magnetization regimes at the two wavelengths.}
\end{figure}

Figure\ \ref{expfit} shows the experimental data from Fig.\ 2 of the main text together with the result of the fits according to Eq.\ (\ref{expeq}).  The results are reported in Tab.\ I of the main text, where the time constants of the Co R-XMLD have been multiplied by factors of 2 and for the amplitudes the square roots of the values according to Eq.\ (\ref{expeq}) are presented, considering the proportionality of the R-XMLD signal to the square of the sublattice magnetization in CoO. 

\subsection*{Modeling of ultrafast dynamics for below-bandgap laser excitation}

\subsubsection*{\textbf{Simulation method}}
To model the dynamics following a \SI{800}{\nano\metre} laser pulse, we use a combination of atomistic spin-dynamics simulations and two-temperature models for each layer.

The energetics of the magnetic degrees of freedom in the FM/AFM bilayer are described by an Heisenberg Hamiltonian,
\begin{small}
\begin{align}\label{equ Hamiltonian}
            \mathcal{H}= -\sum_{ij} J_{ij} \bm{{S_{i}}} \cdot \bm{{S_{j}}},
\end{align}
\end{small}
where $J_{ij}$ are the coupling constants and the $\bm{S}_i$ are unit vectors along the direction of the magnetic moment of the Fe or Co atom located at lattice site $i$. Relativistic effects, which manifest as magneto-crystalline anisotropies or Dzyaloshinskii–Moriya interactions, only lead to small corrections to the Heisenberg term and can thus be safely neglected here.

In order to facilitate the modeling, we make a couple of approximations. First, we replace the actual FM/AFM heterostructure with two simple cubic layers (consisting each of 9 ML), one representing Fe and the other for CoO, with perfect stacking. Second, we restrict the $J_{ij}$ to nearest-neighbor coupling and use values of $J^\mathrm{Fe}= \SI{62.42}{\milli\electronvolt}$ and $J^\mathrm{CoO} =\pm \SI{17.53}{\milli\electronvolt}$ within the respective layers. This way, the simulations reproduce the experimental values for the Curie/N{\'e}el temperatures of $\SI{1043}{\kelvin}$ for Fe and $\SI{293}{\kelvin}$ for CoO in the thermodynamic limit. Note that the sign of the coupling constant in CoO is chosen to positive (favoring parallel alignment) for interactions within each ML and negative (favoring antiparallel alignment) in the perpendicular direction. Consequently, the groundstate of CoO is a layered AFM, same as in the experiments.

Due to the assumption of perfect stacking of the Fe and the CoO layers, the Fe magnetic moments at the interface couple to exactly one Co magnetic moment. The corresponding coupling $J^\mathrm{if}$ is taken as a free parameter.

The dynamics of the magnetic moments are described using the stochastic Landau-Lifshitz-Gilbert (LLG) equation \cite{Landau1935,Gilbert2004,Nowak2007} 
\begin{align}
            & \dot{\bm{S}_i}=-\frac{\gamma_i}{\mu_i}\bm{S}_i \times (\bm{H}_i+\bm{\zeta}_i)+(\alpha^\mathrm{e}_i+\alpha^\mathrm{ph}_i)\bm{S}_i \times \dot{\bm{S}_i} \, , \label{equ LLG}
\end{align}
where $\gamma_i$ is the gyromagnetic ratio, $\mu_i$ the saturation magnetic moment and $\alpha^\mathrm{e(ph)}_i$ the electronic (phononic) Gilbert damping parameter associated with the magnetic moment at lattice site $i$.

The first term in Eq.~\eqref{equ LLG} describes precession in an effective magnetic field $\bm{H}_i = - \partial \mathcal{H} / \partial\bm{S}_i$, which is a result of the interaction with neighboring magnetic moments, and a stochastic field $\bm{\zeta}_i$ modeling the fluctuating torques due to the interaction with phononic and electronic heat baths \cite{brown1963thermal}. The stochastic field has the properties $\langle \bm{\zeta}_i\rangle=0$ and
\begin{align}
     \langle \bm{\zeta}_i(t)\bm{\zeta}_i^\mathrm{T}(t') \rangle = 2\frac{\mu_i}{\gamma_i}k_\mathrm{B}  (\alpha^\mathrm{e}_i T^\mathrm{e}_i + \alpha^\mathrm{ph}_i T^\mathrm{ph}_i)\mathbbm{1} \delta_{ij}\delta(t-t'), \label{equ noise}
\end{align}
where $k_\mathrm{B}$ is the Boltzmann constant and $T^\mathrm{e(ph)}_i$ is the electronic (phononic) temperature at each lattice site. Here, we assume homogeneous electron and phonon temperatures in each layer. 

The second term in Eq.~\eqref{equ LLG} is the so-called Gilbert damping term, which describes dissipation of energy and angular momentum to the heat baths. Due to the large bandgap, there is no coupling of the Co magnetic moments to the electronic heat bath in the CoO layer, i.e.,\ $\alpha^\mathrm{e}_\mathrm{CoO}=0$. Furthermore, we assume $\alpha_\mathrm{Fe}^\mathrm{ph} = 0$, which has been shown to be a reasonable approximation for 3$d$ ferromagnets \cite{Zahn2021,Zahn2022}.

The dynamics of the heat baths are described within a multi-temperature model for the temperatures of Fe electrons $T^\mathrm{e}_\mathrm{Fe}$, Fe phonons $T^\mathrm{ph}_\mathrm{Fe}$ and CoO phonons $T^\mathrm{ph}_\mathrm{CoO}$, which reads
\begin{small}
    \begin{equation}
        \begin{aligned}\label{equ TTM}
            & C_\mathrm{Fe}^\mathrm{e} \dot{T}_\mathrm{Fe}^\mathrm{e} = G_\mathrm{Fe}^\mathrm{e-ph}({T}_\mathrm{Fe}^\mathrm{ph} - {T}_\mathrm{Fe}^\mathrm{e}) - \frac{\partial \mathcal{H}_\mathrm{Fe}}{\partial t} + P_\mathrm{Fe}(t) ,\\
            & C_\mathrm{Fe}^\mathrm{ph} \dot{T}_\mathrm{Fe}^\mathrm{ph} = G_\mathrm{Fe}^\mathrm{e-ph}({T}_\mathrm{Fe}^\mathrm{e} - {T}_\mathrm{Fe}^\mathrm{ph}) +  G_\mathrm{if}^\mathrm{ph-ph}({T}_\mathrm{CoO}^\mathrm{ph} - {T}_\mathrm{Fe}^\mathrm{ph}) ,\\  
            & C_\mathrm{CoO}^\mathrm{ph} \dot{T}_\mathrm{CoO}^\mathrm{ph} = G_\mathrm{if}^\mathrm{ph-ph}({T}_\mathrm{Fe}^\mathrm{ph} - {T}_\mathrm{CoO}^\mathrm{ph}) - \frac{\partial \mathcal{H}_\mathrm{CoO}}{\partial t}
             + G_\mathrm{sink}^\mathrm{ph-ph}({T}_\mathrm{Ag}^\mathrm{ph} - {T}_\mathrm{CoO}^\mathrm{ph}).
        \end{aligned}
    \end{equation}
\end{small}
We use a Sommerfeld approximation for the electron heat capacity, $C_\mathrm{Fe}^\mathrm{e}= \gamma_{Fe}^\mathrm{e} T_\mathrm{Fe}^\mathrm{e}$. The phonon heat capacities $C^\mathrm{ph}$, the phonon-phonon $G^\mathrm{ph-ph}$ and the phonon-electron energy transfer rates  $G^\mathrm{e-ph}$ are assumed to be constant. The terms $\partial\mathcal{H}_\mathrm{Fe}/\partial t$ and $\partial\mathcal{H}_\mathrm{CoO}/\partial t$ account for energy transfer to the magnetic degrees of freedom. Note that the change of (magnetic) interface energy is divided equally between both terms.
The laser power absorbed by the Fe electrons is denoted by $P_\mathrm{Fe}$. The last term in Eq.~\eqref{equ TTM} leads to energy losses of the FM/AFM bilayer to the Ag substrate, which is held at constant temperature ${T}_\mathrm{Ag}^\mathrm{ph}$.

At this stage, the model includes energy transfer between the Fe and CoO layer due to phonons (via $G_\mathrm{if}^\mathrm{ph-ph}$) and magnons (via $J^\mathrm{if}$). Since these transfer channels only become relevant at timescales much longer than the experimentally measured quenching of the AFM order -- a few $\si{\pico\second}$ for magnons and even 10s of $\si{\pico\second}$ for phonons (see below) -- we introduce a third channel of energy transfer by considering a direct coupling between the laser-excited electrons in Fe and the Co magnetic moments, similar to what was done by Wust \textit{et al.}\ \cite{wust2022indirect} for describing laser-induced spin dynamics in Pt/NiO. Here, we model this coupling by introducing an effective Gilbert damping parameter $\alpha_\mathrm{Fe\rightarrow CoO}^\mathrm{e}$ and altering the LLG and the stochastic field for the Co magnetic moments as follows,
\begin{align}
    \dot{\bm{S}_i}&=-\frac{\gamma_i}{\mu_i}\bm{S}_i \times (\bm{H}_i+\bm{\zeta}_i)+(\alpha^\mathrm{e}_{\mathrm{Fe}\rightarrow \mathrm{CoO}}+\alpha^\mathrm{ph}_\mathrm{CoO})\bm{S}_i \times \dot{\bm{S}_i}, 
    \\
    \langle \bm{\zeta}_i(t)\bm{\zeta}_i^\mathrm{T}(t') \rangle &= 2\frac{\mu_i}{\gamma_i}k_\mathrm{B}  (\alpha^\mathrm{e}_{\mathrm{Fe}\rightarrow \mathrm{CoO}} T^\mathrm{e}_\mathrm{Fe} + \alpha^\mathrm{ph}_\mathrm{CoO} T^\mathrm{ph}_\mathrm{CoO})\mathbbm{1} \delta_{ij}\delta(t-t').
\end{align}
This way, the energy transfer from laser-excited hot electrons in Fe to Co magnetic moments in the adjacent layer is described in an incoherent manner, rather than in a coherent one as done previously in Refs.~\cite{Chirac2020,Ritzmann2020b,Weissenhofer2023}. One can also interpret this effect as the inverse of the spin-pumping mechanism from a magnetic system to a paramagnetic conductor  \cite{Tserkovnyak2002}.

Furthermore, we assume that the effective Gilbert damping parameter scales linearly with the temperature of the Fe electrons, $\alpha_{\mathrm{Fe}\rightarrow \mathrm{CoO}}^\mathrm{e}= \xi T_\mathrm{Fe}^\mathrm{e}$, because we expect the associated energy transfer to be strongly peaked during the laser pulse (when the electronic temperature reaches thousands of $\si{\kelvin}$).

The LLG and the multi-temperature model are integrated using Heuns method with a timestep of $\SI{0.5}{\femto\second}$ for a cubic grid with $9$ ML for each of the Fe and CoO layers and $100 \times 100$ magnetic moments per ML for the results shown in the main paper and $40 \times 40$ magnetic moments for the results shown in the supplemental material. We use open boundary conditions along the stacking direction and periodic boundary conditions in the plane perpendicular to it. The absorbed laser fluence is modeled as a Gaussian with a FWHM of \SI{60}{\femto\second}. The total absorbed fluence is taken as a free parameter.

\subsubsection*{\textbf{Model parameters}}
Table \ref{Lit_Parameters} shows the fixed values from the literature that were used for the modeling.  The curves in Fig.\ 2 of the main text were obtained by adjusting the parameters as shown in Tab.\ \ref{Free_Parameters}.  Take note that different parameter settings may be able to 
reproduce the observed behavior, particularly given the noisy nature of the R-XMLD signal.

\begin{table}[h!]
        \begin{adjustbox}{scale = 0.9,center}
        \begin{tabular}{|c|c|c|}
            \hline 
            Parameter & Value & Taken from\\
            \hline
            $J^\mathrm{Fe}$ [meV] & 62.42 & $T_\mathrm{C}$ \\
            \hline
            $J^\mathrm{CoO}$ [meV] & 17.53 & $T_\mathrm{C}$\\
            \hline
            $\gamma^\mathrm{e}_\mathrm{Fe}$ $[\frac{\text{meV}}{\text{K}^\text{2} \text{atom}}]$ & 2.48$\times$10$^{-5}$ & \cite{Ritzmann2020} \\
            \hline
            $C_\mathrm{Fe}^\mathrm{ph}$ $[\frac{\text{meV}}{\text{K atom}}]$ & 0.233  & \cite{Ritzmann2020}\\
            \hline
            $C_\mathrm{CoO}^\mathrm{ph}$ $[\frac{\text{meV}}{\text{K  atom}}]$ & 0.569  & \cite{NISTCoO}\\
            \hline
            $G_\mathrm{Fe}^\mathrm{e-ph}$ $[\frac{\text{meV}}{\text{K atom fs}}]$ & 7.73$\times$10$^{-5}$  & \cite{Ritzmann2020}\\
            \hline
            $\mu_\mathrm{Fe}$ $[\mu_\mathrm{B}]$ & 2.2 & \cite{Kuebler1981}\\
            \hline
            $\mu_\mathrm{CoO}$ $[\mu_\mathrm{B}]$ & 3.8 & \cite{HerrmannRonzaud1978}\\
            \hline
        \end{tabular}
        \end{adjustbox}
        \caption{Fixed parameter values based on literature. The exchange constants we calculated from the Curie temperatures using the relation $1.44J=k_\mathrm{B}T_\mathrm{C}$, which holds true for simple cubic systems with only nearest neighbor exchange \cite{Garanin1996}.} 
        \label{Lit_Parameters}
    \end{table}
    \begin{table}[h!]
        \begin{adjustbox}{scale = 0.9,center}
        \begin{tabular}{|c|c|}
            \hline 
            Parameter & Value\\
            \hline
            $P_\mathrm{Fe}$ $[\frac{\text{meV}}{\text{atom}}]$ & 255\\
            \hline
            $\alpha_\mathrm{Fe}^\mathrm{e}$ & 0.00485\\
            \hline
            $J^\mathrm{if}$ [meV] & 40\\
            \hline
            $\alpha_\mathrm{CoO}^\mathrm{ph}$ & 0.035\\
            \hline
            $\xi$ [$10^{-3} \text{K}^{\text{-1}}$] & 0.00125\\
            \hline
            $G_\mathrm{if}^\mathrm{ph-ph}$ $[\frac{\text{meV}}{\text{atom fs}}]$ & 9.1$\times$10$^{-6}$\\
            \hline
            $G_\mathrm{sink}^\mathrm{ph-ph}$ $[\frac{\text{meV}}{\text{atom fs}}]$ & 1.52$\times$10$^{-4}$\\
            \hline
           \end{tabular}
           \end{adjustbox}
        \caption{Adjusted free parameter values to simulate the experimental data at 800 nm pump.}
        \label{Free_Parameters}
    \end{table}

\subsubsection*{\textbf{Energy transfer by phonons}}
To isolate the contribution of phonons to the energy transfer from Fe to CoO,
the other 
energy transfer
channels were closed 
by using 
$J^\mathrm{if} = 0$ and $\xi = 0$.
For the other parameters, we used the same values as given in Tab.\ \ref{Lit_Parameters} and Tab.\ \ref{Free_Parameters}.

\begin{figure}[h!]
\includegraphics[width=0.5\columnwidth]{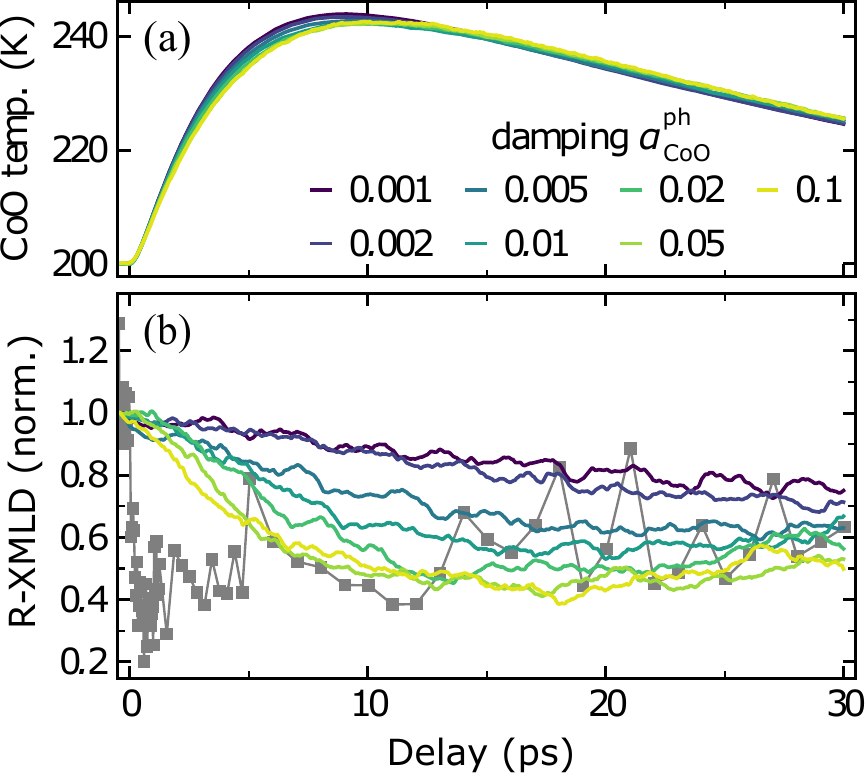}
\caption{\label{phononic_contribution}(a) The CoO lattice temperatures corresponding to (b) the simulated Co R-XMLD for different damping parameters due to CoO phonons, $\alpha^\mathrm{ph}_\mathrm{CoO}$. Considering only the phononic contribution effectively slows down the CoO demagnetization to tens of picoseconds.}
\end{figure}

By varying $\alpha_\mathrm{CoO}^\mathrm{ph}$, which 
has a significant impact on
how quickly the CoO spin order (squared to get R-XMLD) adjusts to variations in phonon temperature, it is apparent that CoO demagnetization by phonons occurs at timescales of tens of picoseconds even for large values of $\alpha_\mathrm{CoO}^\mathrm{ph}$, as shown in Fig.\ \ref{phononic_contribution}. The phonon temperatures only change on this timescale and, therefore, limit the demagnetization rate via phonons. 
 
Higher values of $G_\mathrm{if}^\mathrm{ph-ph}$ would cause faster temperature rise of the CoO phonons, but would also result in quicker remagnetization of Fe than the experimental observation. Hence, it suffices to say that indirect energy transfer through the phonon channel cannot explain the subpicosecond demagnetization of CoO.

\subsubsection*{\textbf{Energy transfer by magnons}}

\begin{figure}[h!]
\includegraphics[width=0.95\columnwidth]{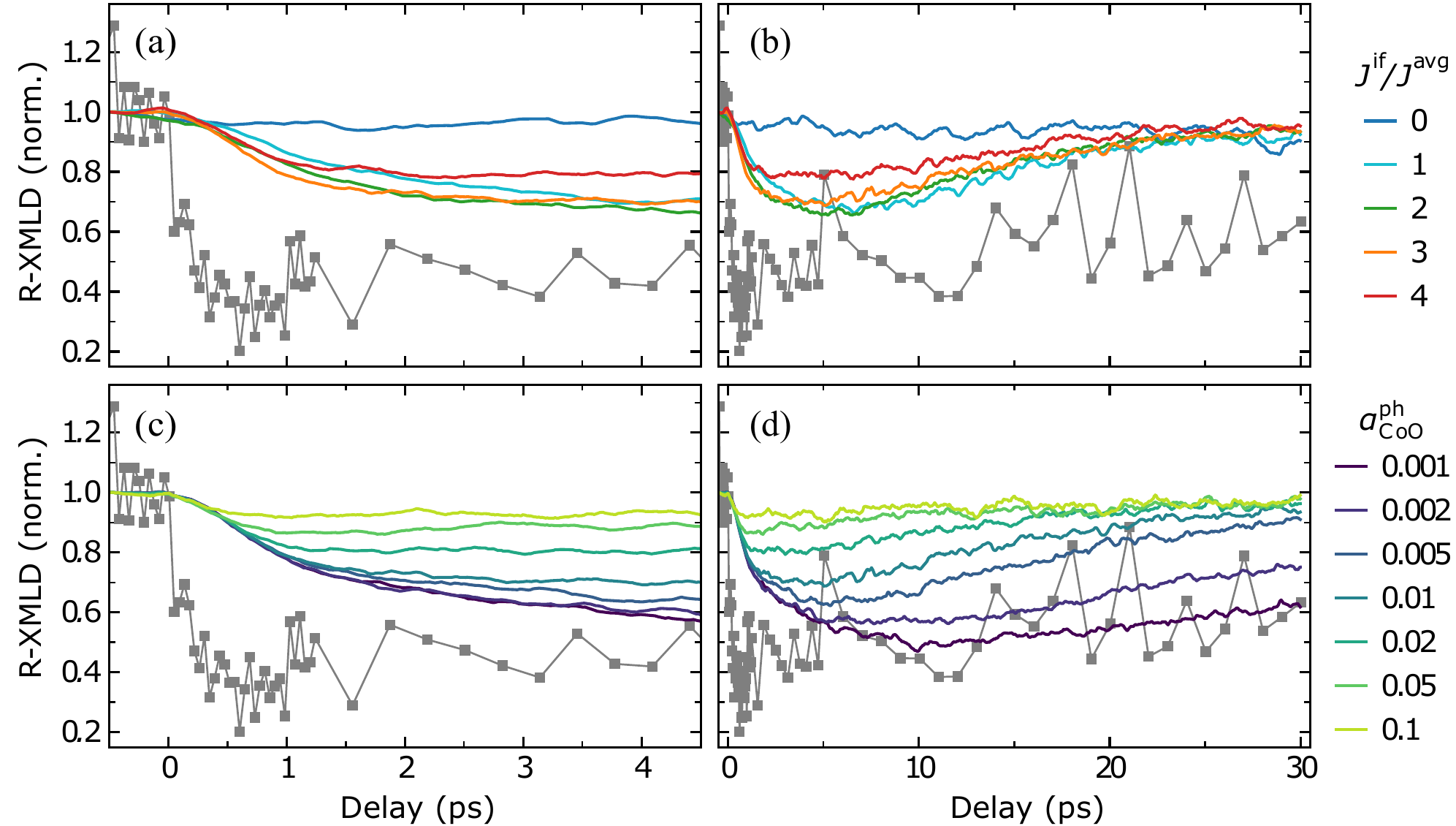}
\caption{\label{magnonic_contribution}(a) and (b): Simulated R-XMLD for different $J^\mathrm{if}$ 
in units of $J^\mathrm{avg} = \langle J^\mathrm{Fe} + J^\mathrm{CoO} \rangle$, with a fixed $\alpha_\mathrm{CoO}^\mathrm{ph} = 0.01$, at short and long delay-time range, respectively. (c) and (d): Simulated R-XMLD for different $\alpha_\mathrm{CoO}^\mathrm{ph}$, with a fixed $J^\mathrm{if} = 3J^\mathrm{avg}$, at short and long delay-time range on the left and right, respectively. 
Considering only
the magnonic contribution effectively slows the CoO demagnetization down to several picoseconds.}
\end{figure}

The exchange interaction $J^\mathrm{if}$ describes the magnonic contribution to the demagnetization of the CoO sublattice. Its influence is extracted by blocking the electronic and phononic channels,
achieved by setting
$ \xi = 0$ and $G_\mathrm{sink}^\mathrm{ph-ph} = \infty$, respectively (the latter 
ensures that all energy transferred to the CoO phonons is immediately lost to the sink
, the Ag substrate, rather than accumulating in the CoO spin system).
In the top panels of Fig.\ \ref{magnonic_contribution}, for a given value of $\alpha_\mathrm{CoO}^\mathrm{ph} = 0.01$, the interfacial exchange coupling $J^\mathrm{if}$ 
was varied. In the bottom panel, the spin-phonon coupling in CoO, $\alpha_\mathrm{ph}^\mathrm{CoO}$, was varied, keeping $J^\mathrm{if} = 3J^\mathrm{avg}$ constant. 
Based on our theory calculations, it is evident that magnetic interactions at the interface lead to a demagnetization of CoO,
but only over a period of several picoseconds. 

\subsubsection*{\textbf{Energy transfer by electrons}}

The impact of Fe electron coupling to CoO spins for magnetization dynamics was isolated in the simulations by switching off the other channels
through setting
$J^\mathrm{if} = 0$ and $G_\mathrm{sink}^\mathrm{ph-ph} = \infty$. 

\begin{figure}[h!]
\includegraphics[width=1\columnwidth]{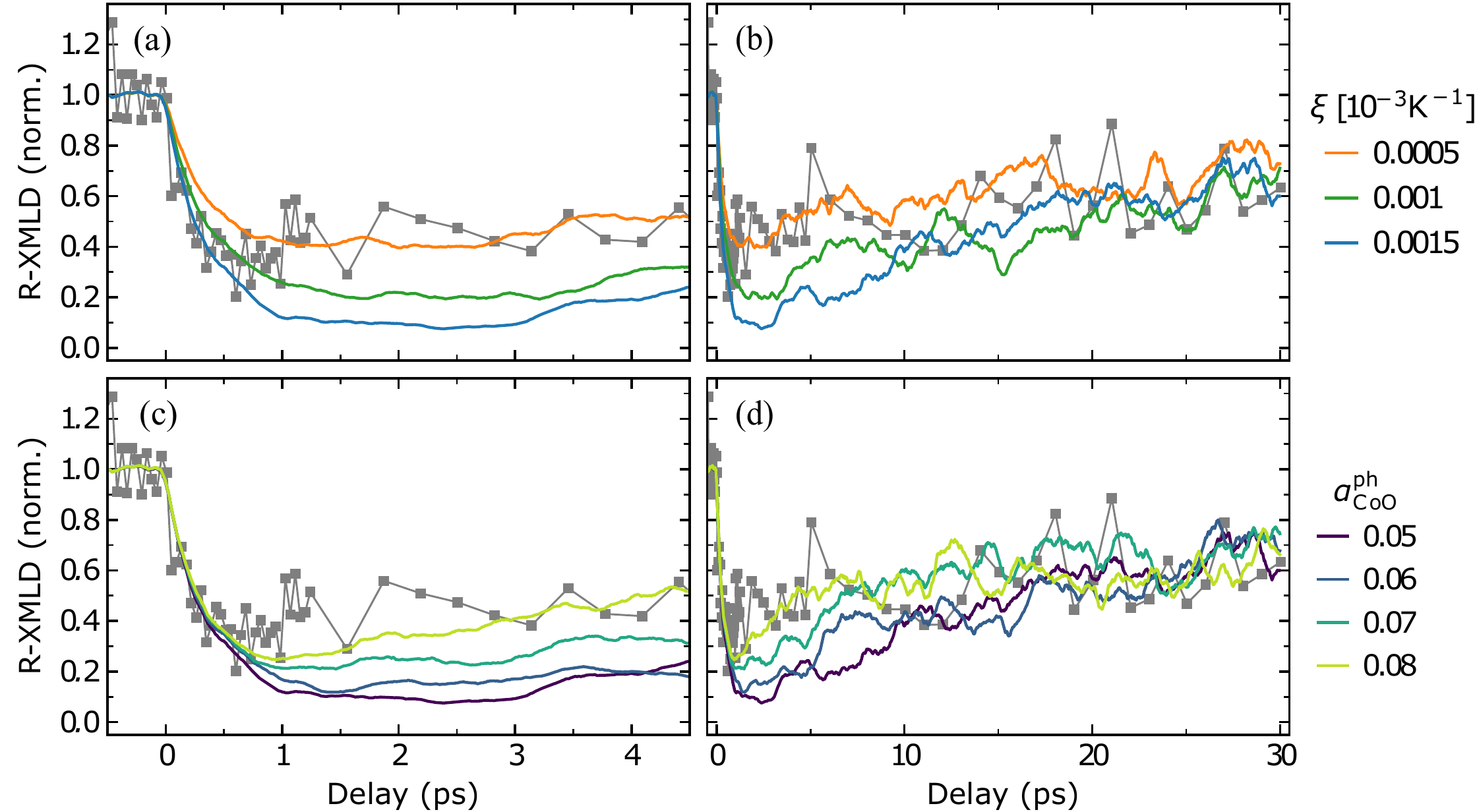}
\caption{\label{electronic_contribution}(a) and (b) Simulated R-XMLD for different 
$\xi$, with a fixed $\alpha_\mathrm{CoO}^\mathrm{ph} = 0.05$, in short and long delay range, respectively. (c) and (d) Simulated R-XMLD for different $\alpha_\mathrm{CoO}^\mathrm{ph}$, with a fixed 
$\xi = \SI{0.00125e-3}{\kelvin^{-1}}$, in short and long delay range, respectively. 
The electronic contribution to the energy transfer induces CoO demagnetization at subpicosecond timescales.}
\end{figure}

The top panels of Fig.\ \ref{electronic_contribution} illustrate how 
$\xi$
affects the change in dynamics for fixed $\alpha_\mathrm{CoO}^\mathrm{ph} = 0.05$, while the bottom panel displays the effect of $\alpha_\mathrm{CoO}^\mathrm{ph}$ for fixed
$\xi$.
As can be seen, this results in the demagnetization of CoO in less than 1 ps. This is because CoO spins interact directly with electrons in Fe and follow their temperature (with a delay specified by the relevant Gilbert damping factor).  Out of the three possible channels of energy transfer from the Fe to the CoO layer, only this one is able to reproduce the experimentally observed ultrafast quenching of the AFM order in CoO.

\subsection*{Modeling of ultrafast dynamics for above-bandgap laser excitation}

\subsubsection*{\textbf{Simulations excluding direct CoO excitation}}

\begin{figure}[!h]
\includegraphics[width=0.95\columnwidth]{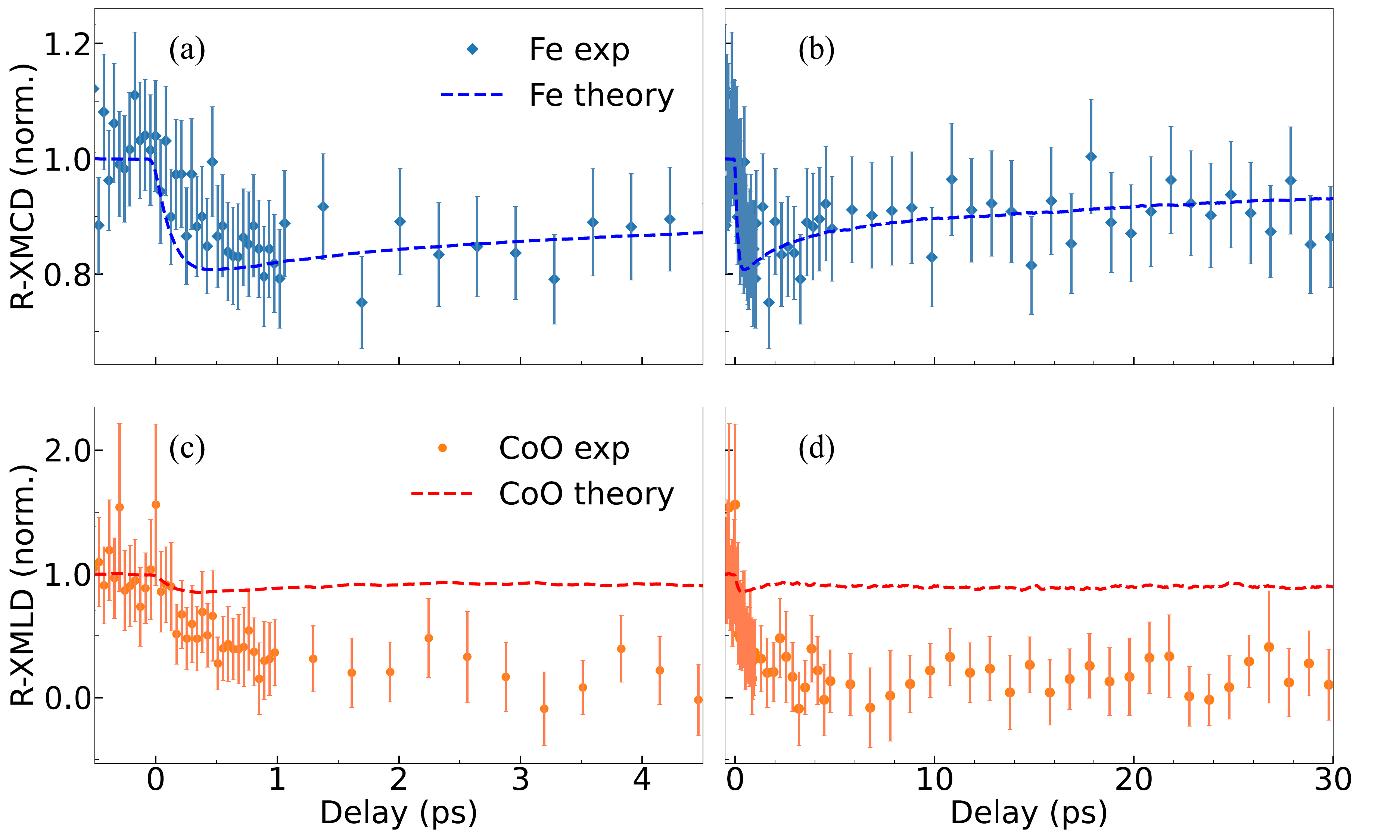}
\caption{\label{400_model_1}
Comparison of experimental data for 400 nm with simulations without direct electronic excitation of CoO.
The top panels show the Fe R-XMCD signal at short 
(a)
and long 
(b)
delay-time ranges.
The bottom panels present the CoO R-XMLD signal at short 
(c)
and long 
(d)
delay-time ranges. The scattered points are the experimental data and the dashed lines are the simulation. 
}
\end{figure}

As a first step, we
utilized the exact same model and the same parameters acquired as for the 800 nm data, taking into account all three possible indirect energy transfer channels. We simply adjusted the absorbed fluence in Fe to match the experimentally observed demagnetization amplitude for Fe. The result is shown in Fig.\ \ref{400_model_1}. 

The demagnetization of Fe can be fairly well reproduced.  
However, according to the simulation, there would be significantly less demagnetization of CoO than in the experiment (decrease of R-XMLD signal by only around 10 - 15\% as opposed to the observed approximately 80\%).  This shows that direct optical excitation of the CoO layer has to be responsible for the major part of the ultrafast quenching of AFM order in CoO.
\vspace{-3cm}
\subsubsection*{\textbf{Simulations with direct CoO excitation}
}
In an effort to reproduce the experimental
data at 400 nm pump, we extend the temperature model from Eq.\ (\ref{equ TTM}) by considering also direct energy transfer to CoO electrons upon pumping with 400 nm photons, expressed by $P_\mathrm{CoO}$, and introducing the temperature of the electrons in CoO ${T}_\mathrm{CoO}^\mathrm{e}$ as an additional variable.
Moreover, we introduce a coupling between the CoO electrons and magnetic moments by using a finite Gilbert damping parameter $\alpha_\mathrm{CoO}^\mathrm{e}$ in the stochastic LLG and the stochastic field, see Eqs.~\eqref{equ LLG} and \eqref{equ noise}. 
The equations governing the time evolution of the relevant temperatures then read
\begin{small}
    \begin{equation}
        \begin{aligned}\label{equ TTM400}
            & C_\mathrm{Fe}^\mathrm{e} \dot{T}_\mathrm{Fe}^\mathrm{e} = G_\mathrm{Fe}^\mathrm{e-ph}({T}_\mathrm{Fe}^\mathrm{ph} - {T}_\mathrm{Fe}^\mathrm{e}) - \frac{\partial \mathcal{H}_\mathrm{Fe}}{\partial t} + P_\mathrm{Fe}(t) \, , \\
            & C_\mathrm{Fe}^\mathrm{ph} \dot{T}_\mathrm{Fe}^\mathrm{ph} = G_\mathrm{Fe}^\mathrm{e-ph}({T}_\mathrm{Fe}^\mathrm{e} - {T}_\mathrm{Fe}^\mathrm{ph}) +  G_\mathrm{if}^\mathrm{ph-ph}({T}_\mathrm{CoO}^\mathrm{ph} - {T}_\mathrm{Fe}^\mathrm{ph}) \, , \\
            & C_\mathrm{CoO}^\mathrm{e} \dot{T}_\mathrm{CoO}^\mathrm{e} = G^\mathrm{e-ph}_\mathrm{CoO}(T^\mathrm{ph}_\mathrm{CoO}- T^\mathrm{e}_\mathrm{CoO}) - \frac{\alpha^\mathrm{e}_\mathrm{CoO}}{\alpha^\mathrm{e}_\mathrm{CoO}+\alpha^\mathrm{ph}_\mathrm{CoO}}\frac{\partial \mathcal{H}_\mathrm{CoO}}{\partial t} + P_\mathrm{CoO}(t) \, , \\
            & C_\mathrm{CoO}^\mathrm{ph} \dot{T}_\mathrm{CoO}^\mathrm{ph} = G^\mathrm{e-ph}_\mathrm{CoO}(T^\mathrm{e}_\mathrm{CoO}- T^\mathrm{ph}_\mathrm{CoO}) + G_\mathrm{if}^\mathrm{ph-ph}({T}_\mathrm{Fe}^\mathrm{ph} - {T}_\mathrm{CoO}^\mathrm{ph}) - \frac{\alpha^\mathrm{ph}_\mathrm{CoO}}{\alpha^\mathrm{e}_\mathrm{CoO}+\alpha^\mathrm{ph}_\mathrm{CoO}}\frac{\partial \mathcal{H}_\mathrm{CoO}}{\partial t}\\
            & ~~~~~~~~~~~~~~~~~ + G_\mathrm{sink}^\mathrm{ph-ph}({T}_\mathrm{Ag}^\mathrm{ph} - {T}_\mathrm{CoO}^\mathrm{ph}) \,.
        \end{aligned}
    \end{equation}
\end{small}
Here, we have introduced an electron-specific heat capacity of CoO, $C_\mathrm{CoO}^\mathrm{e}$,
which -- for simplicity -- we assume to be linear in temperature, i.e., $C_\mathrm{CoO}^\mathrm{e}= \gamma_\mathrm{CoO}^\mathrm{e} {T}_\mathrm{CoO}^\mathrm{e}$.
Note that we have constructed the model such that the change in magnetic energy in CoO $\partial \mathcal{H}_\mathrm{CoO}/\partial t$ is distributed to the electronic and phononic heat bath according to the strength of coupling, i.e., of the associated Gilbert damping parameter.
In comparison to the 800-nm model of Eq.\ (\ref{equ TTM}), there are thus four additional free parameters: $\gamma_\mathrm{CoO}^\mathrm{e}$, $G_\mathrm{CoO}^\mathrm{e-ph}$, $\alpha_\mathrm{CoO}^\mathrm{e}$, and $P_\mathrm{CoO}$. For simplicity, we ignored any additional coupling terms with the CoO electrons other than $G_\mathrm{CoO}^\mathrm{e-ph}$ and $\alpha_\mathrm{CoO}^\mathrm{e}$.

To reproduce the experimental data for 400 nm pump in Fig.\ 2 of the main text, we used the same set of parameters as in Tab.\ \ref{Lit_Parameters} and Tab.\ \ref{Free_Parameters}, except for $P_\mathrm{Fe} = 85$ meV/atom, with the four additional parameters as listed below in Tab.\ \ref{400_Parameters}.

    \begin{table}[th!]
        \begin{adjustbox}{scale = 0.9,center}
        \begin{tabular}{|c|c|}
            \hline 
            Parameter & Values\\
            \hline
            $\gamma^\mathrm{e}_\mathrm{Fe}$ $[\frac{\text{meV}}{\text{K}^\text{2} \text{atom}}]$ & 1.44$\times$10$^{-4}$\\
            \hline
            $G_\mathrm{CoO}^\mathrm{e-ph}$ $[\frac{\text{meV}}{\text{K atom fs}}]$ & 1.52$\times$10$^{-4}$\\
            \hline
            $\alpha_\mathrm{CoO}^\mathrm{e}$ & 0.003\\
            \hline
            $P_\mathrm{CoO}$ $[\frac{\text{meV}}{\text{atom}}]$ & 110\\
            \hline
        \end{tabular}
        \end{adjustbox}
        \caption{Additional parameters for the simulation of the experimental data for 400 nm pump.}
        \label{400_Parameters}
    \end{table}

\providecommand{\noopsort}[1]{}\providecommand{\singleletter}[1]{#1}%